\documentclass[journal]{IEEEtran}
\usepackage{amsmath,amsfonts,amssymb}
\usepackage{algorithmic}
\usepackage{array}
\usepackage[caption=false,font=normalsize,labelfont=sf,textfont=sf]{subfig}

\usepackage{textcomp}
\usepackage{stfloats}
\usepackage{url}
\usepackage{verbatim}
\usepackage{graphicx}
\def\BibTeX{{\rm B\kern-.05em{\sc i\kern-.025em b}\kern-.08em
    T\kern-.1667em\lower.7ex\hbox{E}\kern-.125emX}}
\usepackage{balance}

\usepackage{flushend}
\usepackage{cite}
\usepackage{xcolor}
\usepackage{multirow}
\usepackage{arydshln}
\usepackage{tikz,pgfplots}
\usepackage{tikz-3dplot}
\tikzset{>=latex}
\usepgfplotslibrary{groupplots}
\usetikzlibrary{calc}
\pgfdeclarelayer{background}
\pgfdeclarelayer{foreground}
\pgfsetlayers{background,main,foreground} 
\usetikzlibrary{shapes}
\usetikzlibrary {shapes.multipart}
\usetikzlibrary{positioning}
\usepgfplotslibrary{fillbetween}
\usepackage{tabularx}
\newcolumntype{Y}{>{\centering\arraybackslash}X}
\newcolumntype{L}{>{\raggedleft\arraybackslash}X}
\newcolumntype{R}{>{\raggedright\arraybackslash}X}
\usepackage{cite}

\usepackage[normalem]{ulem}
\usepackage{color}

\newcommand\pro[1]{\mbox{#1\%}} 
\newcommand\pros[1]{\mbox{#1\%}}

\usepackage{breakurl}
 
\pgfplotsset{compat=1.16,
  }
\usepackage{listings}
\lstdefinestyle{mystyle}{
    commentstyle=\color{codegreen},
    keywordstyle=\color{magenta},
    numberstyle=\tiny\color{codegray},
    stringstyle=\color{codepurple},
    basicstyle=\ttfamily\footnotesize,
    breakatwhitespace=true,
    breaklines=true,
    captionpos=b,
    keepspaces=true,
    numbersep=5pt,
    showspaces=false,
    showstringspaces=false,
    showtabs=false,
    tabsize=2,
    lineskip=0.5pt,
    belowskip=1.0pt}
\definecolor{background}{rgb}{0.925,0.925,0.925}
\definecolor{codegreen}{rgb}{0,0.6,0}
\definecolor{codegray}{rgb}{0.5,0.5,0.5}
\definecolor{codepurple}{rgb}{0.58,0,0.82}
\usepackage{makecell}
\usepackage{circuitikz}
\DeclareMathSymbol{\shortminus}{\mathbin}{AMSa}{"39}

\newcommand\clean[1]{{\textcolor{red}{}}}

\begin{document}
\bstctlcite{IEEEexample:BSTcontrol}

\title{ Energy Demand Prediction for Hardware Video Decoders Using Software Profiling}                                      
\author{\IEEEauthorblockN{	Matthias Kr\"anzler, Christian Herglotz, and Andr\'e Kaup}\\
\IEEEauthorblockA{Multimedia Communications and Signal Processing,\\
Friedrich-Alexander-Universität Erlangen-Nürnberg (FAU),
Cauerstr. 7, 91058 Erlangen, Germany\\
Email: \{matthias.kraenzler, christian.herglotz, and andre.kaup\}@fau.de}
\thanks{Manuscript received December 12, 2024. Corresponding author: Matthias Kr\"anzler.
The authors are with the Chair of Multimedia Communications and Signal Processing, Friedrich-Alexander-Universit\"at Erlangen-N\"urnberg (FAU), Erlangen, Germany (e-mail: matthias.kraenzler@fau.de; andre.kaup@fau.de; christian.herglotz@fau.de). }}

\markboth{IEEE Journal on Consumer Electronics}%
{Journal}

\maketitle
\begin{abstract}
    Energy efficiency for video communications is essential for mobile devices with a limited battery capacity. Therefore, hardware decoder implementations are commonly used to significantly reduce the energetic load of video playback. The energy consumption of such a hardware implementation largely depends on a previously published specification of a video coding standard that defines which coding tools and methods are included. However, during the standardization of a video coding standard, the energy demand of a hardware implementation is unknown. Hence, the hardware complexity of coding tools is judged subjectively by experts from the field of hardware programming without using standardized assessment procedures. To solve this problem, we propose a method that accurately models the energy demand of existing hardware decoders with an average error of \pro{1.79} by exploiting information from software decoder profiling. Motivated by the low estimation error, we propose a hardware decoding energy metric that can predict and estimate the energy demand of an unknown hardware implementation using information from existing hardware decoder implementations and available software implementations of the future video decoder. By using multiple video coding standards for model training, we can predict the relative energy demand of an unknown hardware decoder with a minimum error of \pro{4.54} without using the corresponding hardware decoder for training. 
\end{abstract}


\section{Introduction}
\IEEEPARstart{M}{obile} communications connect people worldwide over the Internet and are significant contributors to everyone's entertainment. 
In recent years, it was observed that the mobile IP data traffic increased significantly. 
According to \cite{Ericson2023}, the total mobile IP data traffic increased from 39.5 EB/month in 2019 to 154.7 EB/month in 2023. 
In 2028, the data traffic will reach 472 EB/month, corresponding to an increase of over ten times within nine years. 
This rise is mainly caused by the higher demand for video content, which contributed around \pro{71} in 2023 and will increase to over \pro{80} by 2028.

Simultaneously, it is estimated that \pro{1} of the global greenhouse gas emissions (GHG) were caused by video communications in 2018 \cite{Efoui-Hess2019}.
Therefore, the rise in video communications also increases GHG emissions due to a higher watch time and data traffic. 
Consequently, it is crucial to improve energy efficiency of video communications to reduce GHG emissions \cite{Herglotz2023c}. 
Furthermore, it is also important to consider that the consumption of video content on mobile devices leads to a significant reduction in battery lifetime \cite {Herglotz19a}.

\begin{figure}[!t]
    \definecolor{gray}{HTML}{CCCCCC}
    \resizebox{\columnwidth}{!}{%
    \tdplotsetmaincoords{70}{110}
    \begin{tikzpicture}[scale=0.4, tdplot_main_coords]
        
        \coordinate (O) at (0,0,0);
        \draw[thick,->,red,line width = 0.1cm] (0,0,0) -- (15,0,0) node[anchor=north]{Hardware energy consumption?};
        \draw[thick,->,line width = 0.1cm] (0,0,0) -- (0,12,0) node[anchor=north]{Bit rate in Mbps};
        \draw[thick,->,line width = 0.1cm] (0,0,0) -- (0,0,12) node[anchor=south]{PSNR in dB};
        \path[fill=gray,draw=none]
        (0,0.625,2) -- (0,1.25,4) -- (0,2.5,6) -- (0,5,8) -- (0,10,10) --  (0,5,10) -- (0,2.5,8) -- (0,1.25,6) -- (0,0.625,4) -- (0,0.3125,2) --  (0,0.625,2);
        \draw[thick,black,line width = 0.05cm] (0,5,8) -- (0,10,10) node[anchor=south]{Codec A};
        \draw[thick,black,line width = 0.05cm] (0,2.5,6) -- (0,5,8) node[anchor=south]{};
        \draw[thick,black,line width = 0.05cm] (0,1.25,4) -- (0,2.5,6) node[anchor=south]{};
        \draw[thick,black,line width = 0.05cm] (0,0.625,2) -- (0,1.25,4) node[anchor=south]{};
        \draw[thick,black,line width = 0.05cm] (0,0.3125,1) -- (0,0.625,2) node[anchor=south]{};
    
       \draw[thick,blue,line width = 0.05cm] (0,5,10) -- (0,10,12) node[anchor=south]{Codec B}; 
       \draw[thick,blue,line width = 0.05cm] (0,2.5,8) -- (0,5,10) node[anchor=south]{};
        \draw[thick,blue,line width = 0.05cm] (0,1.25,6) -- (0,2.5,8) node[anchor=south]{};
        \draw[thick,blue,line width = 0.05cm] (0,0.625,4) -- (0,1.25,6) node[anchor=south]{};
        \draw[thick,blue,line width = 0.05cm] (0,0.3125,2) -- (0,0.625,4) node[anchor=south]{};
        \node at (0,7.5,6) {BD-Rate-PSNR};

        \draw[thick,black,line width = 0.025cm] (0,7.2,6.5) -- (0,4,8.25) node[anchor=south]{};
    
    \end{tikzpicture}
    }

\vspace*{-0.3cm}
\caption{Illustration of the challenges during standardization, which optimizes the compression efficiency of future video coding standards. However, the hardware energy consumption is typically unknown during the development of the standard.}

\vspace*{-0.5cm}
\label{fig:Problem}
\end{figure}
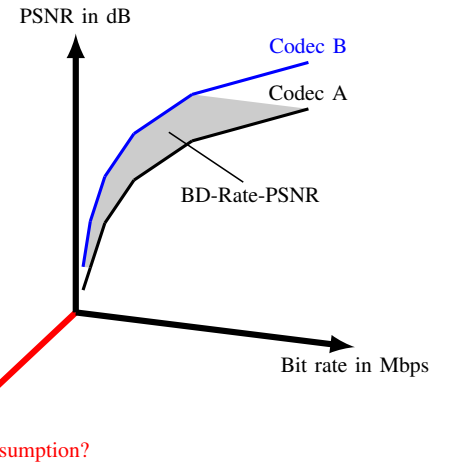

To address these challenges, standardization bodies such as ISO, ITU, or the Alliance for Open Media (AOM) have focuesed on reducing the global video data traffic by enhancing compression efficiency through new video coding standards. 
AOM was founded in 2015 and released its first video coding standard AOMedia Video 1 (AV1) in 2018.
The goal was an improved compression efficiency compared to standards such as VP9, High-Efficiency Video Coding (HEVC) or Advanced Video Coding (AVC). 
AOM is now developing AOM Video Model (AVM) as a successor to AV1 \cite{avm}.
Similarly, ISO and ITU established the Joint Video Exploration Team (JVET) in 2015 with the same goal as AOM.
In 2020, their video coding standard Versatile Video Coding (VVC) was published.

Although both AV1 and VVC demonstrated significant improvements in compression efficiency, they require substantially higher energy for video decoding. 

According to \cite{Kraenzler2020MMSP}, the energy demand of VVC for software (SW) decoding is over \pro{80} higher than that of HEVC. Additionally, an optimized AV1 video decoder was shown to consume \pro{16} more energy than a comparable VP9 decoder \cite{Kranzler_2024}. For HEVC, the additional energy demand was found to be \pro{18} higher compared to VP9.

However, for hardware (HW) decoders, AV1 increases energy demand by over \pro{115} compared to VP9, while HEVC decoding shows a modest increase of \pro{6}. This disparity highlights a significant mismatch in energy efficiency between HW and SW decoders, creating a critical challenge for future hardware implementations, where minimizing energy demand is especially important for mobile devices that typically rely on HW decoder implementations.

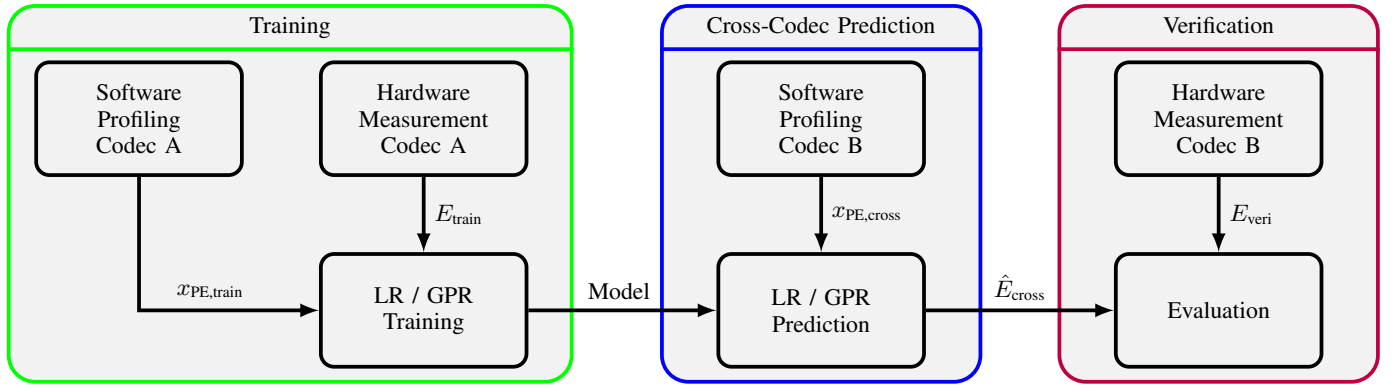
\begin{figure*}[!t]
    \vspace{-0.5cm}
        \tikzstyle{every node}=[font=\small]
    
        \tikzset{line/.style={draw, thick, -latex'}}
        \tikzset{box/.style={draw, rectangle, rounded corners=5pt, thick, node distance=1cm, text width=2.5cm,line width=0.05cm,text centered, minimum height=1.5cm}}
        \tikzset{superbox/.style={draw, rectangle,  rounded corners=10pt, thick, node distance=0.5cm, text width=6cm, text centered,text depth=4.25cm, minimum height = 5cm,fill=black!5, line width=0.05cm}}

        \tikzset{container/.style={draw, rectangle, dashed, inner sep=1cm}}
        \tikzset{line/.style={draw, line width=0.05cm, -latex}}
        \tikzset{evalline/.style={draw,dashed, line width=0.05cm, latex-latex,black!50}}
    
          \begin{tikzpicture}
            \node [box] (swp1) {Software\\Profiling\\Codec A};
            \node [box, right=1cm of swp1] (hw1) {Hardware\\Measurement\\Codec A};
            \node [box, below=1cm of hw1] (gprt) {LR / GPR\\Training};
            \node [box, right=2.5cm of hw1] (swp2) {Software\\Profiling\\Codec B};
            \node [box, below=1cm of swp2] (gpre) {LR / GPR\\Prediction};
            \node [box, right=2.5cm of swp2] (hw2) {Hardware\\Measurement\\Codec B};
            \node [box, below=1cm of hw2] (ls) {Evaluation};
    
            \begin{pgfonlayer}{background}
                \node [superbox,above=-4.3cm of swp2, text width=4cm,draw=blue,text depth=4.5cm] (Standardization) {Cross-Codec Prediction};
                \node [superbox, left=1.15cm of Standardization, text width=7.25cm,draw=green,text depth=4.5cm] (Training) {Training};
                \node [superbox, above=-4.3cm of hw2, text width=4cm,draw=purple,text depth=4.5cm] (Verification) {Verification};
    
                \coordinate (lineStart1) at ($(Standardization.north west)+(0,-0.6cm)$);
                \coordinate (lineEnd1) at ($(Standardization.north east)+(0,-0.6cm)$);
                \draw[thick,line width=0.05cm,blue] (lineStart1) -- (lineEnd1);
    
                \coordinate (lineStart2) at ($(Training.north west)+(0,-0.6cm)$);
                \coordinate (lineEnd2) at ($(Training.north east)+(0,-0.6cm)$);
                \draw[thick,line width=0.05cm,green] (lineStart2) -- (lineEnd2);
    
                \coordinate (lineStart3) at ($(Verification.north west)+(0,-0.6cm)$);
                \coordinate (lineEnd3) at ($(Verification.north east)+(0,-0.6cm)$);
                \draw[thick,line width=0.05cm,purple] (lineStart3) -- (lineEnd3);
            \end{pgfonlayer}
    
            \path [line] (swp1) |- (gprt) node[midway,above right] {\hspace{0.25cm} $x_{\textrm{PE,train}}$};
    
            \path [line] ($(hw1.south)+(-0,0cm)$) -- ($(gprt.north)+(-0,0cm)$) node[midway, right] { $E_{\textrm{train}}$};
    
            \path [line] (swp2) -- (gpre) node[midway, right] { $x_{\textrm{PE,cross}}$};
            \path [line] (gprt) -- (gpre) node[midway, above] {\hspace{-0.2cm} Model};
            \path [line] (gpre) -- (ls) node[midway, above] { $\hat{E}_{\textrm{cross}}$};
            
            \path [line] ($(hw2.south)+(-0,0cm)$) -- ($(ls.north)+(-0,0cm)$) node[midway, right] { $E_{\textrm{veri}}$};
    
        \end{tikzpicture}
    
      \vspace*{-0.2cm}
    \caption{Overview of the workflow to predict the energy demand of an unknown HW video decoder. In total, there are three modules with training, cross-codec prediction, and verification.}
    
    \vspace*{-0.5cm}
    \label{fig:SolutionOverview}
    \end{figure*}

Figure~\ref{fig:Problem} shows the problem that is faced during the development of a new video coding standard. 
Usually, the standardization committee compares a state-of-the-art video coding standard (Codec A) with the newly proposed coding standard (Codec B). 
Then, the bit rate savings of codec B over codec A are evaluated in terms of Bj{\o}ntegaard Delta Bit Rate (BDR) \cite{VCEG-M33,Herglotz2023}. 
However, the HW energy demand and chip area cannot be evaluated during the standardization, because HW implementations are not yet available. 
Thus, tradeoffs between the compression efficiency and the energy efficiency cannot be evaluated. 
A worst-case scenario could involve a HW implementation with high power demand, limiting the support of high video resolutions on mobile devices. Furthermore, a rapid battery drain could negatively impact the user’s quality of experience.

An overview of the proposed methodology to predict the energy demand of an unknown HW video decoder implementation is shown in Fig.~\ref{fig:SolutionOverview}. 
This workflow illustrates the systematic guideline to predict the energy demand of a HW decoder by using SW decoder profiling, which involves the modules training, cross-codec prediction, and verification. 
We select coding standards with an available SW and HW implementation (Codec A) for the training of the energy models (green box in Fig.~\ref{fig:SolutionOverview}). 
First, the energy demand of a given HW decoder is measured ($E_{\textrm{train}}$). 
Then, the SW decoders are analyzed with instruction profilers such as Perf or Valgrind \cite{valgrind}.
Thereby, we derive the occurrences of processor events (PEs) such as instructions, memory accesses, branches, and cache misses (c.f., $x_{\textrm{PE}}$).

For the cross-codec prediction module (blue box in Fig.~\ref{fig:SolutionOverview}), we leverage the pre-trained energy models from the training module to predict the energy demand in the context of a different codec, Codec B. Under the standardization scenario, it is assumed that only a SW implementation is available. By performing SW profiling on this implementation, we extract the PEs, which are then used to estimate the HW decoder energy demand, $\hat{E}_{\textrm{cross}}$.

Finally, as indicated by the red box in Fig.~\ref{fig:SolutionOverview}, we evaluate the accuracy of the energy models. To this end, we use a coding standard with an available HW implementation that differs from the training coding standard. This evaluation determines whether cross-codec prediction provides meaningful and reliable energy estimates for the HW decoder of Codec B, and also assesses whether cross-codec prediction is suitable for use in standardization. AV1 will always be used for verification throughout the paper.

In this work, we will first present a literature review of previous work on analyzing, modeling, and optimizing video decoder energy demand in Section~\ref{sec:Literature}. 
After that, we introduce the evaluation setup and define our metrics in Section~\ref{sec:Setup}. 
In Section~\ref{sec:Regression}, we will discuss the regression methods and models used to train the energy demand of the video decoders. 
In Section~\ref{sec:SoftwareModeling}, the results of the SW decoder energy demand modeling is shown.
In Section~\ref{sec:HardwareModeling}, we discuss how the energy demand of HW decoders can be modeled using SW decoders. 
Then, in Section~\ref{sec:HardwarePrediction}, we will show how the energy of an unknown video decoder HW design can be predicted with existing SW decoder implementations.
Additionally, in Section~\ref{sec:Application}, we show an application of this framework in standardization, where it can be used to estimate the expected hardware energy demand.
We conclude the paper in Section~\ref{sec:Conclusion} with a summary of the findings and an outlook on future research.

\section{Literature Review}
\label{sec:Literature}
\noindent
The complexity of video coding standards is commonly analyzed in terms of energy, time, and chip area, followed by modeling experimental results to understand and optimize the complexity. These models are used for further optimization.

\subsubsection{Analysis}
\noindent
The complexity of video coding standards has been studied in multiple works. \cite{Laude_2019} examined HEVC, AVC, VVC, and AV1 in terms of encoding time, decoding time, and compression efficiency. \cite{Kraenzler2020MMSP} analyzed the energy demand of HEVC and VVC decoders, proposing optimization methods for energy reduction. Other works compared VVC with HEVC \cite{Mercat_2021} and AV1 \cite{Nguyen_2021}. \cite{Kranzler_2024} evaluated the energy efficiency of AVC, HEVC, VP9, VVC, AV1, and AVM across both HW and SW decoders. \cite{Katsenou_2022} proposed a metric that relates energy demand and bit rate for AV1, VVC, VP9, and HEVC, enabling a tradeoff evaluation across multiple video coding standards.

HW decoder implementations for AV1 and VVC face challenges in implementing new coding tools, as noted by \cite{Correa_2021}. Further, hardware designs for VVC coding tools such as adaptive loop filters and transformations were proposed in \cite{Hamidouche_2022, Farhat_2021, Farhat_2022}. \cite{Saldanha_2020} assessed both video coding standards, showing a need for efficient algorithms and complexity reductions for new coding tools. Similarly, \cite{Viitanen2022} developed a VVC intra encoder based on HEVC and evaluated the complexity of new coding tools. \cite{Wieckowski2022} evaluated the increased encoding complexity of VVC over HEVC and proposed optimization strategies to limit the VVC encoder search space. 

\subsubsection{Modeling}
\noindent
The energy demand of SW decoders was modeled using processor events (PEs) of the CPU, as in \cite{Herglotz2017}, or through bitstream features, as proposed in \cite{MallikarachchiTalagalaH.EtAl2017, HerglotzSpringerReichenbachEtAl2018, Kraenzler2019, Kraenzler2020ICIP}. Typically, these models provide high estimation accuracy for SW decoders. The HW decoder energy demand has been modeled using high-level parameters, such as video resolution \cite{Herglotz_2018}. In \cite{HerglotzCoulombeVazquezEtAl2020}, the power demand of mobile devices during video streaming was accurately described using parameters like display brightness, frame rate, and bit rate, with these factors determining the majority of the energy demand. In \cite{Kraenzler2023}, we presented a method to model HW decoder energy demand using SW profiling, but the estimation error was over \pro{10}, a limitation we aim to improve in this work.

\subsubsection{Optimization}
\noindent
Several optimization methods have been explored to reduce energy consumption in video coding. \cite{HerglotzHeindelKaup} used bitstream features to train a model for the SW decoding energy demand, extending rate-distortion optimization for HEVC encoding to include decoding energy demand, leading to significant reductions. \cite{Kraenzler21, Kraenzler2022, Kranzler_2022} focused on design space exploration for VVC, evaluating coding tools based on their tradeoff between bit rate and energy efficiency. This optimization led to reductions in energy demand of up to \pro{50} compared to default VVC bitstreams. \cite{Tissier_2019, Choi_2022} explored optimization opportunities for VVC intra encoders. In \cite{Amirpour_2023}, preset selection was optimized for HEVC encoders to balance encoding time, energy demand, and compression efficiency.

In our prior work \cite{Kraenzler2023}, we combined the concepts from \cite{Herglotz2017} and \cite{Herglotz_2018} to create a unified solution for modeling HW decoders using SW profiling. While \cite{Herglotz2017} introduced SW decoder energy profiling with Valgrind and \cite{Herglotz_2018} focused on HW energy demand modeling with bitstream features, our approach resulted in a higher estimation error of \pro{13.14} for HW energy modeling using the Valgrind 13PE model. This paper addresses that issue by employing Gaussian Process Regression (GPR) to reduce the estimation error. Additionally, we use a different HW decoder, as the one used in \cite{Kraenzler2023} required video playback via HDMI, introducing errors due to activities not captured by SW decoder profiling.

A key novelty of this paper is the introduction of a cross-codec prediction framework. Unlike previous works that focused on estimating energy demand for a single codec, this work proposes a methodology for predicting the energy demand of an unknown HW video decoder using profiling data from multiple video coding standards.

\begin{table}[t!]
    \def\arraystretch{1.1}
    \caption{Overview of encoder and decoder SW implementations used for each video coding standard.}

    \vspace*{-0.3cm}
    \label{tab:SoftwareCoDecs}
    \begin{center}
    \begin{tabular}{ l | l | l | l } 
         &  Encoder &  Reference Decoder\!\! &  Optimized Decoder \\
         \hline\hline
    AVC\!\!  & x264 \cite{x264} & JM \cite{JM} & FFmpeg \cite{ffmpeg} \\
    \hdashline
    HEVC\!\! & x265 \cite{x265}\!\! & HM \cite{HM} & openHEVC \cite{openHEVC}\!\!\!\! \\
    \hdashline
    VVC\!\!  & VVenC \cite{VVenCSoftware}\!\! & VTM \cite{VTM}\! & VVdeC \cite{VVdeCSoftware} \\
    \hdashline
    VP9\!\!  & libvpx \cite{libvpx}\!\! & libvpx & FFmpeg \\
    \hdashline
    AV1\!\!  & libaom \cite{libaom}\!\! & libaom & dav1d \cite{dav1d} \\
    \hdashline
    AVM\!\!  & avm \cite{avm}\!\! & avm & / \\
    \end{tabular}
    \end{center}
    \vspace*{-0.5cm}
\end{table}

\section{Setup and Metrics}
\label{sec:Setup}

\subsection{Video Encoder and Decoder Setup}
\label{subsec:VideoCodecs}
\noindent 
In this work, we evaluate six video coding standards, as listed in Table~\ref{tab:SoftwareCoDecs}. For each standard, we use the encoders specified in the second column. Additionally, we evaluate the reference decoder implementation used during the development of the respective video coding standard (third column). Finally, we assess an optimized decoder implementation that is commonly used in practical applications.
As the coding specification of AVM is still under development, there is no available optimized decoder yet.

For the encoding of the video sequences, we follow the common test conditions (CTCs) of AOM \cite{CWGB005oV1} and selected the highest quality preset, ensuring full utilization of available coding tools for both encoding and decoding.
From the CTCs, we take the video sequences of class A1 with 4K resolution, class A2 with full HD resolution, class A3 with HD resolution, and class B with full HD resolution. 
The sequences of class \mbox{A1-3} show natural content, and the sequences of class B show synthetic screen content. 
Each sequence is encoded with the test conditions randomaccess (RA) and lowdelay B (LB) as described in the CTC \cite{CWGB005oV1}. 
The commands for the encoding of the bitstreams and the quantization parameters that were used for each test condition can be found in~\cite{Kranzler_2024}.

\subsection{Measurement Setup}
\label{subsec:Measurement}
\noindent 
To derive the pure decoding energy demand $E_{\textrm{dec}}$, we conduct two alternating measurements as outlined in \cite{HerglotzSpringerReichenbachEtAl2018}. First, the energy demand during the decoding process is measured, followed by the energy demand in an idle state. Finally, $E_{\textrm{dec}}$ is calculated as the difference between these two measurements.
For statistical correctness, we check the measurement series for one bitstream with a
confidence interval test with a maximum deviation of less than \pro{2} from the actual values with a probability of \pro{99} as described in \cite{HerglotzSpringerReichenbachEtAl2018}.

In the following, we distinguish between two measurement setups that are similar to \cite{Kraenzler2023} and are built up as follows:
\subsubsection{Measurement Setup Software (MSS)}
\noindent 
First, a desktop PC is used to measure the energy demand of SW decoders. The PC is equipped with an Intel i7-8700 CPU running CentOS as the operating system. The energy demand is measured using the CPU’s internal power meter, Running Average Power Limit (RAPL) \cite{DavidGorbatovHanebutteEtAl2010}. The advantage of RAPL is that it can be directly accessed by the operating system. \cite{Katsenou2024} demonstrated that RAPL measurements correlate highly with those from an external power meter. All SW decoders are evaluated under single-threaded decoding.

\subsubsection{Measurement Setup Hardware (MSH)}
\noindent
Secondly, the energy demand of HW decoder implementations is measured on the Rock 5 Model B single-board computer (SBC) by Radxa. This SBC features an octa-core ARM processor, comprising a quad-core Cortex-A76 and Cortex-A55 CPU \cite{Rock5B}. The board supports HW decoder implementations for the video coding standards AVC, HEVC, VP9, and AV1. Ubuntu is used as the operating system, and the HW decoder is accessed via FFmpeg. The decoding process occurs without storing or displaying the decoded output. The power demand is measured using the high-precision external power meter LMG611 by ZES Zimmer, connected to the SBC’s main power supply jack.

\section{Linear Regression and Gaussian Process Regression}
\label{sec:Regression}
\noindent 
In the following, we will introduce two methods to train the parameters of the energy models, Linear Regression (LR) and Gaussian process regression (GPR). 

\subsection{Linear Regression}
According to \cite{Izenman2013}, a LR model can be defined as follows:
\begin{equation}
Y = \beta_0 + \sum_{i \in N} \beta_i \cdot X_i  + \epsilon,
\label{eq:LR}
\end{equation}
where $Y$ denotes the output variable to be estimated, $N$ is the number of model parameters, $i$ is the parameter index, $X_i$ represents the input variables, $\beta_i$ are the unknown parameters of the model, and $\epsilon$ is the error term. We assume that the noise follows a normal distribution with a mean of zero and variance $\sigma^2_n$, which is expressed as:
\begin{equation}
\epsilon \sim \mathcal{N}(0, \sigma^2_n).
\label{eq:Noise}
\end{equation}
Thus, a linear model cannot perfectly estimate the output variable $\hat{Y}$ due to the noise factor $\epsilon$ affecting the ground truth measurement of $Y$.

In this work, the output variable of the regression in \eqref{eq:LR} is the decoding energy demand $E_{\textrm{dec}}$. The goal of the regression is to predict or estimate the decoding energy demand $\hat{E}{\textrm{dec}}$ as accurately as possible. Mathematically, the linear regression model for the decoding energy demand is expressed as:
\begin{equation}
\hat{E}_{\textrm{dec}} = \sum_{i \in \textrm{Features}} X_{i} \cdot e_{i},
\label{eq:LinearRegression}
\end{equation}
where $e_i$ is the energy demand coefficient for parameter $i$, which represents the energy required to process one instance of $i$. The energy coefficient $e_i$ corresponds to the parameter $\beta_i$ in \eqref{eq:LR}.

To derive the values of $e_i$, we use a least-squares curve fitting algorithm based on the trust-region-reflective method \cite{ColemanLi1996}. In Section~\ref{subsec:Models}, we introduce the energy models along with their specific features.

\subsection{Gaussian Process Regression}
As an alternative to linear regression (LR), Gaussian Process Regression (GPR), a machine learning-based supervised training regression model, can be utilized \cite{Rasmussen2006}. Unlike the parametric approach of LR in \eqref{eq:LR}, GPR is a non-parametric, kernel-based probabilistic model. It assumes that measurement errors are Gaussian-distributed and centered around the mean of the measurement series (cf. \eqref{eq:Noise}).

According to \cite{Rasmussen2006}, a Gaussian process (GP) is defined by:
\begin{equation}
f(x) \sim \mathcal{GP}\left(m(x), k\left(x_s, x_t\right)\right),
\label{eq:GP}
\end{equation}
where $m(x)$ is the mean function, $x_s$ and $x_t$ are latent variables, and $k(x_s, x_t)$ denotes the covariance (or kernel) function of the GP. For this work, we use an exponential kernel, defined by the following covariance function:
\begin{equation}
k(x_s, x_t) = \sigma^2_f \exp\left(-\frac{|x_s - x_t|}{l}\right) + \sigma^2_n \cdot \delta_{st},
\end{equation}
where $\delta_{st}$ is the Kronecker delta, $l$ is the characteristic length scale, $\sigma^2_f$ is the function variance, and $\sigma^2_n$ is the noise variance. These three parameters are trained as hyperparameters \cite{Rasmussen2006}. The advantage of GPR is its ability to model the variance in the measurement noise (cf. \eqref{eq:Noise}), which is treated as a trainable hyperparameter by the kernel function.

For the mean function $m(x)$, we apply a linear function as in \eqref{eq:LR}. To predict the energy demand with the GPR, we add the kernel function of the GP to the basis functions of the LR model in \eqref{eq:LR} as follows
\begin{equation}
Y = f(x) + h(x)^\intercal \gamma,
\label{eq:GPR}
\end{equation}
where $Y$ represents the decoding energy demand ($E_{\textrm{dec}}$), $f(x)$ is the kernel prediction from \eqref{eq:GP}, and $h(x)^\intercal \gamma$ represents the basis function. This basis function can be expressed as:
\begin{equation}
h(x)^\intercal \gamma = \beta_0 + \sum_{i \in N} \beta_i \cdot X_i = \sum_{i \in \textrm{Features}} X_i \cdot e_i.
\label{eq:GPR}
\end{equation}
Thus, $h(x)$ corresponds to the feature vector $X$ (cf. \eqref{eq:LinearRegression}), and $\gamma$ represents the set of trained feature coefficients $e$ (cf. \eqref{eq:LinearRegression}). In summary, GPR enhances linear regression by adding a kernel function, allowing it to account for uncertainty due to noise.

For the training of the GPR hyperparameters, we use the \textit{fitrgp} function in MATLAB \cite{Matlab2023}. A more detailed description of Gaussian Process Regression can be found in \cite{Rasmussen2006}.

To evaluate the accuracy of our energy model, we use the mean absolute percentage error (MAPE), $\overline{\varepsilon}$, defined as:
\begin{equation}
\overline{\varepsilon} = \frac{1}{M} \sum_{i \in M} \frac{\lvert E_{\textrm{dec},i} - \hat{E}_{\textrm{dec},i} \rvert}{E_{\textrm{dec},i}},
\label{eq:MAPE}
\end{equation}
where $M$ is the number of samples in the dataset, $i$ is the index of the measurement, $E_{\textrm{dec},i}$ is the measured energy demand for sample $i$, and $\hat{E}_{\textrm{dec},i}$ is the corresponding estimated energy demand. A lower value of $\overline{\varepsilon}$ indicates higher accuracy of the energy model.

To prevent overfitting, it is essential to ensure that the data for training and validation are separated for both LR and GPR. Overfitting increases estimation error when applying the model to new datasets \cite{Hastie2009}. Therefore, we employ 10-fold cross-validation for both SW and HW decoding energy modeling.

In addition to GPR, we evaluated other regression methods, such as neural networks and other machine learning techniques from the MATLAB Statistics and Machine Learning Toolbox. However, GPR consistently yielded the most accurate estimation results.

\begin{figure}[!t]

    \begin{center}
    \begin{tikzpicture}

    \tikzstyle{every node}=[font=\small]
    \begin{groupplot}[
         group style={group name=my plots,group size= 1 by 2,vertical sep =1.5cm},
          title style={yshift=-0.25cm},
         height = 10cm,
         width = \textwidth ,set layers,cell picture=true
        ]

        \nextgroupplot[
            width=0.47\textwidth,
            height = 0.3\textwidth,      
            xlabel={SW Decoding Time in s},
            ylabel={SW Energy Demand in J},
            title = {(a) SW Decoder},
           legend cell align = {left},
               xmin=0, xmax=9,
               ymin=0, ymax=230,
               axis lines = left,
               ymajorgrids=true,
               yminorgrids=true,
               xmajorgrids=true,
               xminorgrids=true,
               minor tick num=4,
               minor grid style=dotted,
               grid style = dashed,
            ]
            
            \addplot[only marks,
            color=black,
            fill=white,
             mark=x,
             mark size=3pt,
             line width=1.5pt,
             ]
      coordinates {(2.0066,50.5895) (1.6373,41.2216) (1.3132,33.0838) (1.058,26.5755) (0.88099,22.2062) (0.63115,15.9786) (1.3315,33.6778) (1.1159,28.5611) (0.94883,24.4604) (0.85178,21.8121) (0.74538,19.0621) (0.63197,16.3306) (2.5123,63.7338) (2.1808,55.3607) (1.8453,46.936) (1.5493,39.3325) (1.2944,32.8567) (0.85137,21.3231) (2.1737,55.2305) (1.8025,46) (1.4891,37.7008) (1.2404,31.6248) (1.0315,26.2958) (0.72542,18.2466) (5.4408,137.6888) (3.9986,102.6579) (3.477,88.8502) (3.2575,82.2664) (2.802,70.7949) (2.4156,61.8232) (0.78251,19.9049) (0.64141,16.0449) (0.52837,13.3182) (0.44472,11.0703) (0.37021,9.1471) (0.2601,6.3576) (2.854,72.8728) (2.0081,50.8704) (1.5678,39.7989) (1.2209,31.1433) (0.93456,23.6874) (0.61472,15.3536) (4.7579,122.7325) (4.1546,106.9737) (3.6876,95.7578) (3.3397,85.9056) (3.0668,78.5695) (2.6358,67.2528) (4.6779,118.7382) (3.5627,90.681) (2.8711,73.4272) (2.2896,58.0818) (1.7764,45.0221) (1.0957,27.7529) (3.0508,77.4624) (1.2467,31.0432) (0.71702,18.0148) (0.60865,15.2192) (0.58064,14.5046) (0.49274,12.2143) (0.98473,24.8805) (0.76082,19.283) (0.62349,15.9135) (0.53478,13.5711) (0.45486,11.556) (0.3448,8.6728) (5.1191,127.6103) (4.0796,102.5399) (2.8004,70.538) (1.8057,45.4439) (1.2646,31.9816) (0.83931,21.2139) (5.4251,139.1448) (4.5382,116.5979) (4.0674,104.084) (3.6601,93.6164) (3.3098,84.615) (2.8468,73.2439) (2.2217,56.6254) (1.4895,37.8191) (1.2703,31.9247) (1.0887,27.6219) (0.93396,23.2206) (0.74051,18.5609) (1.4824,37.445) (1.3595,34.2147) (1.1849,29.8734) (1.0295,25.8832) (0.87567,21.9926) (0.61025,15.2081) (0.44433,11.3539) (0.31333,7.8759) (0.26741,6.7738) (0.23823,6.0396) (0.20578,5.1839) (0.1643,4.1374) (0.47319,11.8967) (0.3652,8.999) (0.30827,7.5878) (0.26739,6.5577) (0.23025,5.5785) (0.18752,4.5765) (2.237,56.2877) (1.8251,46.04) (1.4893,37.4278) (1.2391,31.1951) (1.0342,26.067) (0.76139,19.0155) (5.5311,137.7668) (4.7213,118.9621) (3.5112,88.3982) (2.4097,60.5402) (1.6944,42.994) (0.94385,23.9305) (0.99673,25.3859) (0.81114,20.4276) (0.68357,17.0506) (0.56061,14.1044) (0.4823,12.1726) (0.38696,9.6354) (0.54735,13.7314) (0.49972,12.3783) (0.4609,11.5009) (0.42434,10.5216) (0.407,10.0784) (0.36072,8.9185) (5.42,138.6901) (4.6863,118.5441) (3.915,99.7817) (3.2347,82.3108) (2.6076,66.2006) (1.5893,40.3886) (5.077,128.5688) (4.3094,108.6226) (3.4894,88.2009) (2.8676,73.0242) (2.258,57.5949) (1.2987,32.6875) (5.2377,132.448) (4.4877,113.1235) (3.6098,91.6865) (2.89,73.9493) (2.2614,57.8696) (1.3169,33.5287) (8.3982,211.5188) (7.0849,178.9267) (5.8735,148.8639) (4.6819,117.8311) (3.8212,96.9535) (2.8028,71.9272) (5.1855,131.0848) (2.672,67.8715) (1.5998,40.6737) (1.0702,27.1979) (0.90859,23.432) (0.51822,12.8062) (2.1262,53.9624) (1.6935,42.6771) (1.3192,33.2392) (1.1015,27.8694) (0.91376,22.9307) (0.65826,16.3602) (4.9652,126.6072) (4.0846,104.5526) (3.4609,88.7581) (3.0441,77.5521) (2.662,67.7069) (2.2407,57.4811) (1.6658,42.4295) (1.4714,37.2354) (1.2963,32.8785) (1.1277,28.7141) (0.97917,24.7819) (0.75558,19.0623) (2.8745,73.0399) (2.438,62.1637) (2.0366,52.2171) (1.8047,46.0118) (1.5262,38.9551) (1.0534,26.7527) (0.90023,23.0243) (0.74808,18.9709) (0.63209,16.1119) (0.55995,14.0962) (0.49043,12.4298) (0.33767,8.3266) (3.7025,93.3799) (2.7653,70.4728) (2.0662,52.5962) (1.6702,42.3063) (1.3065,33.2451) (0.81446,20.1421) (1.36,34.7719) (1.1993,30.3678) (1.042,26.5849) (0.9049,22.7878) (0.78398,19.7013) (0.57722,14.2383) (5.3843,135.0519) (4.7488,119.6615) (3.8915,98.4309) (3.0643,77.2746) (2.0909,53.3152) (1.02,25.6839) (2.1751,55.0849) (1.7795,45.2683) (1.508,38.4372) (1.3259,33.67) (1.2025,30.6606) (0.91473,22.9442) (1.7984,45.8278) (1.5468,39.395) (1.3836,35.3838) (1.2452,31.9022) (1.1011,28.1952) (0.85843,21.5911) (3.406,86.8632) (2.889,73.036) (2.3379,59.0076) (1.8269,46.0002) (1.3389,33.4126) (0.81512,20.4725) (4.8161,122.3271) (3.8884,99.2966) (3.1422,79.4948) (2.5158,64.0521) (2.201,55.8703) (1.7468,44.3751) (5.6145,143.3486) (4.3822,111.17) (2.8837,73.3693) (1.9832,50.894) (1.5686,39.8339) (1.0967,27.9644) (2.4955,63.4363) (2.099,53.3815) (1.7319,43.5017) (1.4735,37.2645) (1.2741,32.0425) (0.9237,23.215) (1.9532,50.1265) (1.7112,43.5757) (1.4245,36.6071) (1.1992,30.6511) (1.0243,25.9331) (0.75088,18.9475) (1.5009,38.8871) (1.2142,31.0551) (1.0056,25.7329) (0.86667,21.9736) (0.77776,19.8995) (0.6442,16.381) (0.49682,12.7895) (0.38293,9.4779) (0.31943,7.8963) (0.27566,6.7016) (0.24199,5.8582) (0.20479,4.9575)  };

                   \vspace{-1.0cm}

        \nextgroupplot[
            width=0.47\textwidth,
            height = 0.3\textwidth,
            xlabel={Decoding Time in s},
            ylabel={HW Energy Demand in J},
           legend cell align = {left},
              title = {(b) HW Decoder},
               xmin=0, xmax=9,
               ymin=0, ymax=10,
               axis lines = left,
               ymajorgrids=true,
               yminorgrids=true,
               xmajorgrids=true,
               xminorgrids=true,
               minor tick num=4,
               minor grid style=dotted,
               grid style = dashed,
            ]
            
            \addplot[only marks,
            color=black,
            fill=white,
             mark=x,
             mark size=3pt,
             line width=1.5pt,
             ]
      coordinates { (2.0066,2.5064) (1.6373,2.2606) (1.3132,2.0983) (1.058,2.0061) (0.88099,1.9475) (0.63115,1.8409) (1.3315,2.3413) (1.1159,2.2033) (0.94883,2.1205) (0.85178,2.0328) (0.74538,1.9907) (0.63197,1.9286) (2.5123,2.51) (2.1808,2.3559) (1.8453,2.1595) (1.5493,2.0361) (1.2944,2.0077) (0.85137,1.9167) (2.1737,2.7556) (1.8025,2.4757) (1.4891,2.3033) (1.2404,2.2323) (1.0315,2.1643) (0.72542,2.0679) (5.4408,7.2925) (3.9986,6.5816) (3.477,6.4268) (3.2575,6.2958) (2.802,6.2511) (2.4156,6.1906) (0.78251,1.1874) (0.64141,1.1169) (0.52837,1.1072) (0.44472,1.0449) (0.37021,1.0595) (0.2601,1.0614) (2.854,2.8434) (2.0081,2.4244) (1.5678,2.1395) (1.2209,1.9841) (0.93456,1.8503) (0.61472,1.9067) (4.7579,6.6786) (4.1546,6.5142) (3.6876,6.4366) (3.3397,6.3783) (3.0668,6.3411) (2.6358,6.312) (4.6779,3.9526) (3.5627,3.2386) (2.8711,2.8237) (2.2896,2.4557) (1.7764,2.2751) (1.0957,2.0577) (3.0508,3.0858) (1.2467,2.2354) (0.71702,1.9183) (0.60865,1.8504) (0.58064,1.8707) (0.49274,1.8592) (0.98473,1.4037) (0.76082,1.2628) (0.62349,1.1968) (0.53478,1.1419) (0.45486,1.1172) (0.3448,1.0937) (5.1191,4.4683) (4.0796,3.7192) (2.8004,3.0992) (1.8057,2.4815) (1.2646,2.1683) (0.83931,1.9895) (5.4251,7.1877) (4.5382,6.864) (4.0674,6.6859) (3.6601,6.5657) (3.3098,6.4867) (2.8468,6.4362) (2.2217,2.4937) (1.4895,2.1674) (1.2703,2.0855) (1.0887,2.0118) (0.93396,1.9562) (0.74051,1.893) (1.4824,2.2494) (1.3595,2.1678) (1.1849,2.042) (1.0295,1.9431) (0.87567,1.8672) (0.61025,1.7724) (0.44433,1.1086) (0.31333,1.0305) (0.26741,1.0175) (0.23823,1.0484) (0.20578,1.0292) (0.1643,1.0392) (0.47319,1.116) (0.3652,1.0511) (0.30827,1.0431) (0.26739,1.0491) (0.23025,1.0548) (0.18752,1.0551) (2.237,2.7092) (1.8251,2.4574) (1.4893,2.2515) (1.2391,2.113) (1.0342,2.0844) (0.76139,2.0245) (5.5311,2.0625) (4.7213,1.9939) (3.5112,1.9168) (2.4097,1.8857) (1.6944,1.8991) (0.94385,1.9225) (0.99673,3.909) (0.81114,3.4593) (0.68357,2.9912) (0.56061,2.7283) (0.4823,2.352) (0.38696,2.0103) (0.54735,3.9736) (0.49972,3.4482) (0.4609,3.0448) (0.42434,2.6819) (0.407,2.2152) (0.36072,1.846) (5.42,7.8558) (4.6863,6.5463) (3.915,6.1273) (3.2347,5.9535) (2.6076,5.9658) (1.5893,5.8102) (5.077,9.3945) (4.3094,8.3153) (3.4894,7.7041) (2.8676,7.2678) (2.258,7.0128) (1.2987,6.6807) (5.2377,3.7985) (4.4877,2.2449) (3.6098,2.0063) (2.89,1.9725) (2.2614,1.9313) (1.3169,1.9081) (8.3982,2.1661) (7.0849,2.0367) (5.8735,1.92) (4.6819,1.8538) (3.8212,1.7786) (2.8028,1.8275) (5.1855,6.8545) (2.672,6.5374) (1.5998,6.3357) (1.0702,6.2701) (0.90859,6.1547) (0.51822,6.0083) (2.1262,2.1472) (1.6935,2.0426) (1.3192,1.9907) (1.1015,1.9231) (0.91376,1.9441) (0.65826,1.9066) (4.9652,2.7161) (4.0846,2.3144) (3.4609,2.0196) (3.0441,1.8501) (2.662,1.7531) (2.2407,1.7457) (1.6658,1.2643) (1.4714,1.1995) (1.2963,1.1288) (1.1277,1.0904) (0.97917,1.0822) (0.75558,1.049) (2.8745,3.6295) (2.438,2.7613) (2.0366,2.3163) (1.8047,2.1805) (1.5262,2.0746) (1.0534,1.9955) (0.90023,1.8775) (0.74808,1.8665) (0.63209,1.8534) (0.55995,1.8121) (0.49043,1.793) (0.33767,1.7134) (3.7025,4.5551) (2.7653,3.9849) (2.0662,3.4304) (1.6702,3.1944) (1.3065,2.505) (0.81446,1.9544) (1.36,1.9835) (1.1993,1.7493) (1.042,1.6734) (0.9049,1.6591) (0.78398,1.7007) (0.57722,1.7615) (5.3843,3.312) (4.7488,2.9997) (3.8915,2.7033) (3.0643,2.4445) (2.0909,2.1919) (1.02,2.0021) (2.1751,7.4207) (1.7795,6.6621) (1.508,6.5533) (1.3259,6.4844) (1.2025,6.4051) (0.91473,6.2893) (1.7984,7.3351) (1.5468,6.9862) (1.3836,6.6652) (1.2452,6.4557) (1.1011,6.5471) (0.85843,6.3888) (3.406,4.0636) (2.889,3.4669) (2.3379,3.0478) (1.8269,2.563) (1.3389,2.208) (0.81512,2.0183) (4.8161,4.614) (3.8884,3.7383) (3.1422,2.9828) (2.5158,2.4884) (2.201,2.2385) (1.7468,2.0112) (5.6145,1.1835) (4.3822,1.1067) (2.8837,1.0649) (1.9832,1.0578) (1.5686,1.0642) (1.0967,1.049) (2.4955,1.1902) (2.099,1.084) (1.7319,1.0603) (1.4735,1.0581) (1.2741,1.0474) (0.9237,1.0621) (1.9532,1.4373) (1.7112,1.2663) (1.4245,1.1659) (1.1992,1.1048) (1.0243,1.045) (0.75088,1.0375) (1.5009,2.7565) (1.2142,2.4196) (1.0056,2.2203) (0.86667,2.0871) (0.77776,2.097) (0.6442,1.9667) (0.49682,2.0657) (0.38293,1.9966) (0.31943,1.9377) (0.27566,1.9011) (0.24199,1.9401) (0.20479,1.8569)   };
      \addlegendentry{$t_{\mathrm{dec,SW}}$}

      \addplot[only marks,
      color=red,
      fill=black,
       mark=o,
       mark size=3pt,
       line width=0.75pt,
       ]
        coordinates { (0.94984,2.5064) (0.84558,2.2606) (0.75723,2.0983) (0.71739,2.0061) (0.6963,1.9475) (0.65757,1.8409) (0.8819,2.3413) (0.81074,2.2033) (0.762,2.1205) (0.72288,2.0328) (0.69921,1.9907) (0.68473,1.9286) (0.98686,2.51) (0.87801,2.3559) (0.81295,2.1595) (0.75819,2.0361) (0.7382,2.0077) (0.719,1.9167) (1.0388,2.7556) (0.92113,2.4757) (0.83613,2.3033) (0.7728,2.2323) (0.755,2.1643) (0.7268,2.0679) (2.4516,7.2925) (2.1284,6.5816) (2.0308,6.4268) (1.9872,6.2958) (1.9484,6.2511) (1.9303,6.1906) (0.51578,1.1874) (0.47588,1.1169) (0.44315,1.1072) (0.43707,1.0449) (0.43259,1.0595) (0.43,1.0614) (1.2866,2.8434) (0.97662,2.4244) (0.84473,2.1395) (0.74883,1.9841) (0.70475,1.8503) (0.6776,1.9067) (2.1992,6.6786) (2.056,6.5142) (2.0248,6.4366) (2.0092,6.3783) (1.9992,6.3411) (1.9832,6.312) (1.9244,3.9526) (1.4171,3.2386) (1.1088,2.8237) (0.92432,2.4557) (0.82857,2.2751) (0.71786,2.0577) (1.1605,3.0858) (0.80867,2.2354) (0.6856,1.9183) (0.66591,1.8504) (0.66701,1.8707) (0.66419,1.8592) (0.58763,1.4037) (0.52765,1.2628) (0.48134,1.1968) (0.46156,1.1419) (0.45175,1.1172) (0.44664,1.0937) (2.2234,4.4683) (1.6955,3.7192) (1.1745,3.0992) (0.926,2.4815) (0.7844,2.1683) (0.70821,1.9895) (2.4128,7.1877) (2.214,6.864) (2.1412,6.6859) (2.0869,6.5657) (2.0503,6.4867) (2.0156,6.4362) (0.92467,2.4937) (0.78213,2.1674) (0.72385,2.0855) (0.69847,2.0118) (0.69055,1.9562) (0.67852,1.893) (0.81817,2.2494) (0.77517,2.1678) (0.7329,2.042) (0.69421,1.9431) (0.67306,1.8672) (0.644,1.7724) (0.44953,1.1086) (0.42375,1.0305) (0.41722,1.0175) (0.41886,1.0484) (0.41772,1.0292) (0.4275,1.0392) (0.4392,1.116) (0.41958,1.0511) (0.41891,1.0431) (0.42189,1.0491) (0.42546,1.0548) (0.426,1.0551) (0.9884,2.7092) (0.87701,2.4574) (0.79033,2.2515) (0.74118,2.113) (0.71534,2.0844) (0.69507,2.0245) (0.73918,2.0625) (0.698,1.9939) (0.67267,1.9168) (0.66815,1.8857) (0.6755,1.8991) (0.67527,1.9225) (2.1194,3.909) (1.6977,3.4593) (1.2942,2.9912) (1.0414,2.7283) (0.8586,2.352) (0.71684,2.0103) (2.272,3.9736) (1.7868,3.4482) (1.3195,3.0448) (1.0152,2.6819) (0.82,2.2152) (0.67286,1.846) (2.7308,7.8558) (2.3048,6.5463) (2.0308,6.1273) (1.8836,5.9535) (1.8616,5.9658) (1.8138,5.8102) (2.9209,9.3945) (2.6836,8.3153) (2.5116,7.7041) (2.34,7.2678) (2.2324,7.0128) (2.1036,6.6807) (1.6613,3.7985) (0.84587,2.2449) (0.73667,2.0063) (0.6902,1.9725) (0.68294,1.9313) (0.676,1.9081) (0.76818,2.1661) (0.69778,2.0367) (0.66906,1.92) (0.65229,1.8538) (0.63934,1.7786) (0.646,1.8275) (2.2648,6.8545) (2.0832,6.5374) (2.0048,6.3357) (1.9476,6.2701) (1.8836,6.1547) (1.8468,6.0083) (0.7508,2.1472) (0.71225,2.0426) (0.6874,1.9907) (0.67055,1.9231) (0.68556,1.9441) (0.67025,1.9066) (1.2373,2.7161) (0.91846,2.3144) (0.733,2.0196) (0.671,1.8501) (0.63112,1.7531) (0.63506,1.7457) (0.50891,1.2643) (0.482,1.1995) (0.4505,1.1288) (0.439,1.0904) (0.43077,1.0822) (0.4195,1.049) (1.4948,3.6295) (1.0193,2.7613) (0.82893,2.3163) (0.7648,2.1805) (0.73667,2.0746) (0.6956,1.9955) (0.6754,1.8775) (0.66292,1.8665) (0.66086,1.8534) (0.65873,1.8121) (0.64433,1.793) (0.628,1.7134) (2.4436,4.5551) (1.9848,3.9849) (1.5677,3.4304) (1.2317,3.1944) (0.92364,2.505) (0.69574,1.9544) (0.7512,1.9835) (0.668,1.7493) (0.63077,1.6734) (0.6264,1.6591) (0.63455,1.7007) (0.6368,1.7615) (1.264,3.312) (1.088,2.9997) (0.95534,2.7033) (0.86325,2.4445) (0.76375,2.1919) (0.69374,2.0021) (2.5332,7.4207) (2.1636,6.6621) (2.0512,6.5533) (2.012,6.4844) (1.9936,6.4051) (1.9544,6.2893) (2.481,7.3351) (2.264,6.9862) (2.118,6.6652) (2.0252,6.4557) (2.0412,6.5471) (1.9964,6.3888) (2.0919,4.0636) (1.5871,3.4669) (1.1496,3.0478) (0.9072,2.563) (0.7645,2.208) (0.68734,2.0183) (2.4775,4.614) (1.8253,3.7383) (1.2239,2.9828) (0.97219,2.4884) (0.84566,2.2385) (0.7324,2.0112) (0.47228,1.1835) (0.4455,1.1067) (0.43221,1.0649) (0.41308,1.0578) (0.42117,1.0642) (0.43277,1.049) (0.472,1.1902) (0.44415,1.084) (0.4276,1.0603) (0.4144,1.0581) (0.41833,1.0474) (0.42633,1.0621) (0.6505,1.4373) (0.556,1.2663) (0.48013,1.1659) (0.44128,1.1048) (0.42917,1.045) (0.404,1.0375) (1.0214,2.7565) (0.8772,2.4196) (0.78272,2.2203) (0.74347,2.0871) (0.72417,2.097) (0.68157,1.9667) (0.74056,2.0657) (0.7058,1.9966) (0.68174,1.9377) (0.67086,1.9011) (0.67629,1.9401) (0.66182,1.8569) };
      \addlegendentry{$t_{\mathrm{dec,HW}}$}

    \end{groupplot}
    
    \end{tikzpicture}
    \end{center}
    \vspace{-0.6cm}

    \caption{Evaluation of the energy demand for SW Decoding (a) and HW Decoding (b). In (a), the horizontal axis represents the SW decoding time of dav1d. In (b), a distinction is made between SW decoding time (black markers) and HW decoding time (red markers).}
    \label{fig:HW_SW_Plot_DecTime}

    \vspace{-0.6cm}
    \end{figure}
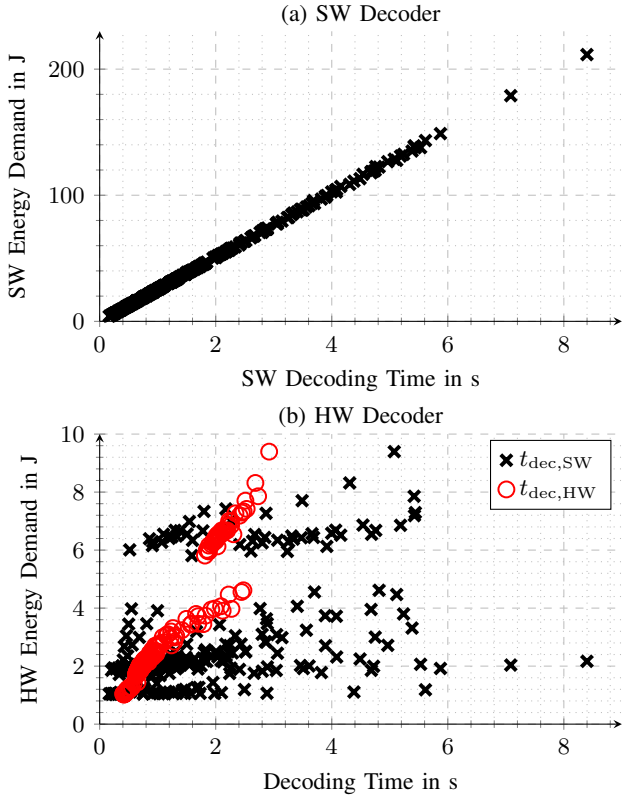

\subsection{Energy Models}
\label{subsec:Models}

For the modeling, we will evaluate the following three parameter sets:

\subsubsection{Temporal Energy Model}
\noindent
For the Temporal model, we utilize the SW decoding time $t_{\textrm{dec,SW}}$ to estimate and predict the SW or HW decoding energy demand $\hat{E}_{\textrm{dec}}$ as follows
\begin{equation}
    \label{eq:temporal}
    \hat{E}_{\textrm{dec}} = t_{\textrm{dec,SW}} \cdot P  + E_{\textrm{offset}}.
\end{equation}
where $P$ is the constant power draw and $E_{\textrm{offset}}$ is the energy offset of the decoding process. 
Based on the assumption that decoding time has a strong correlation with decoding energy demand, $P$ is treated as a linear constant factor.

For standardization, SW decoding time is commonly used to describe the complexity of a video decoder. Consequently, we benchmark our proposed energy models against the Temporal model, the state-of-the-art method for estimating video decoder energy demand.

Figure~\ref{fig:HW_SW_Plot_DecTime} presents an analysis of decoding energy demand for both SW (a) and HW (b) decoders. In both subfigures, black markers represent SW decoding time, while red markers represent HW decoding time.

In Figure~\ref{fig:HW_SW_Plot_DecTime}a, a strong linear relationship between decoding time and energy is evident for SW decoders. However, in Figure~\ref{fig:HW_SW_Plot_DecTime}b, the HW decoder’s energy demand shows a weaker correlation with HW decoding time (red markers). Furthermore, the SW decoding time (black markers) proves inadequate for estimating HW decoding energy, as the markers are widely scattered. Thus, we conclude that SW decoding time is insufficient for accurately estimating HW decoding energy. 

\subsubsection{Perf Energy Model}
\noindent
As recommended by the AOM CTC, we utilize Perf to model the SW decoder energy demand. Perf is a Linux performance monitoring tool that reports various PEs, such as instruction counts and memory reads, derived from the CPU’s hardware counters. It evaluates performance and identifies bottlenecks with minimal temporal overhead. AOM suggests reporting the instruction count, cycle count, and user time for newly proposed coding tools~\cite{CWGB005oV1}. Following this recommendation, we utilize these three features in the Perf CTC model, as described in \cite{Kraenzler2023}.

\subsubsection{Valgrind Energy Model}
\noindent
In addition to the first two models, we employ Valgrind \cite{valgrind} to describe the decoding process through PEs that represent the required CPU instructions. Valgrind simulates a CPU with two cache levels: first-level (L1) and last-level cache (LLC). This work evaluates the Valgrind 13PE model proposed in \cite{Kraenzler2023}, where detailed explanations of the models 13 PEs can be found.

The parameters of the Valgrind model are derived using the following command:
\begin{lstlisting}
    valgrind --tool=callgrind --simulate-cache=yes --dump-instr=yes --collect --jumps=yes --branch=yes --I1=32768,8,64 --D1=32768,8,64 --LL=12582912,24,64 [decoder command]
\end{lstlisting}

Specifying \texttt{I1} (L1 instruction cache), \texttt{D1} (L1 data cache), and \texttt{LL} (LLC) ensures reproducible results across different machines running Valgrind. These parameters reflect the CPU configuration of the desktop PC: the first value specifies the cache size, the second defines the associativity, and the third sets the cache line size.

\begin{table}[!t]

    \def\arraystretch{1.2}
	\centering
	\caption{MAPE results of the SW decoder energy demand modeling. The lowest $\overline{\varepsilon}$ for each decoder is given in bold.}
\vspace{-0.4cm}
    \subfloat[Linear Regression]{
	\begin{tabular}{c l | r  |  r | r }
		\multicolumn{2}{c|}{}  & \multicolumn{1}{c|}{}	& \multicolumn{1}{c|}{~~Perf~~}	& \multicolumn{1}{c}{Valgrind}	 \\
		\multicolumn{2}{c|}{Software Decoder}  & \multicolumn{1}{c|}{Temporal}	& \multicolumn{1}{c|}{ ~~CTC~~}	& \multicolumn{1}{c}{13PE} \\
	\hline \hline
    \multirow{2}{*}{AVC} & JM  & 2.10\% & 5.24\% & \textbf{1.50\%} \\ 
	& FFmpeg  & 5.78\% & \textbf{4.45\%} & 4.47\% \\ 
	\hline
	\multirow{2}{*}{HEVC} & HM  & \textbf{0.68\%} & 2.38\% & 0.87\% \\ 
	& openHEVC  & \textbf{0.73\%} & 1.46\%  & 1.28\% \\ 
	\hline
	\multirow{2}{*}{VP9} & libvpx  & \textbf{0.99\%} & 2.58\% & 1.04\% \\ 
	& FFmpeg  & \textbf{0.93\%} & 1.70\% & 1.26\% \\ 
	\hline
	\multirow{2}{*}{AV1} & libaom  & \textbf{0.96\%} & 1.54\%  & 1.04\% \\ 
	& dav1d  & \textbf{0.87\%} & 1.23\% & 1.36\% \\ 
	\hline
	\multirow{2}{*}{VVC} & VTM  & \textbf{0.71\%} & 3.00\% & 1.63\% \\ 
	& VVdeC  & 1.68\% & 2.54\%  & \textbf{1.49\%} \\ 
	\hline
	\multirow{1}{*}{AVM} & avm  & \textbf{1.29\%} & 3.13\% & 1.61\% \\ 
	\hline
	\hline
	\multicolumn{2}{c|}{Average} & \textbf{1.52\%} & 2.66\% & 1.60\% \\ 
	\end{tabular}
	\label{tab:SoftwareDecoderModelingLR}
    }
\hspace{1.2cm}
\subfloat[Gaussian Process Regression]{
	\begin{tabular}{c l | r  | r  |  r }
		\multicolumn{2}{c|}{}  & \multicolumn{1}{c|}{}	& \multicolumn{1}{c|}{~~Perf~~}	& \multicolumn{1}{c}{Valgrind}	 \\
		\multicolumn{2}{c|}{Software Decoder}  & \multicolumn{1}{c|}{Temporal}	& \multicolumn{1}{c|}{ ~~CTC~~}	& \multicolumn{1}{c}{13PE} \\
	\hline \hline
    \multirow{2}{*}{AVC} & JM  & 2.12\% & 2.14\% & \textbf{1.48\%} \\ 
	                     & FFmpeg  & 6.40\% & \textbf{4.76\%} & 4.98\% \\ 
		\hline
        \multirow{2}{*}{HEVC} & HM  & 0.68\% & 0.97\% &  \textbf{0.63\%} \\ 
	                          & openHEVC  & 0.86\% & 0.95\% & \textbf{0.77\%} \\ 
	\hline
	\multirow{2}{*}{VP9} & libvpx  & 1.33\% & 1.72\% & \textbf{0.77\%} \\ 
	                     & FFmpeg  & 1.18\% & 1.67\% &  \textbf{0.84\%} \\ 
	\hline
	\multirow{2}{*}{AV1} & libaom  & 1.45\% & 1.54\% &  \textbf{0.88\%} \\ 
	                     & dav1d  & 1.05\% & 1.28\% &  \textbf{0.93\%} \\ 
	\hline
	\multirow{2}{*}{VVC} & VTM  & \textbf{0.85\%} & 1.77\% & 1.59\% \\ 
	                     & VVdeC  & 1.62\% & 2.02\% & \textbf{1.06\%} \\ 
	\hline
	\multirow{1}{*}{AVM} & avm  & 1.53\% & 2.00\% & \textbf{1.28\%} \\ 
	\hline
	\hline
    \multicolumn{2}{c|}{Average} & 1.73\% & 1.89\% & \textbf{1.38\%} \\ 
	\end{tabular}
	\label{tab:SoftwareDecoderModelingGPR}
}
\label{tab:SoftwareDecoderModeling}

\vspace{-0.4cm}
\end{table}

\section{Software Energy Modeling}
\label{sec:SoftwareModeling}
\noindent 
In this section, we evaluate the energy models introduced earlier for SW video decoders. Table~\ref{tab:SoftwareDecoderModeling} presents the MAPE for the LR and the GPR. The first column lists the evaluated SW decoders, with results averaged across all decoders in the final row. The evaluated models include Temporal, Perf CTC, and Valgrind 13PE from Section~\ref{sec:Regression}.

The MAPE values in Table~\ref{tab:SoftwareDecoderModeling}a indicate that all SW decoders can be modeled with an average error below \pro{3}. Among the models, the Temporal model achieves the lowest estimation error (\pro{1.52}), followed by the Valgrind 13PE model (\pro{1.60}). These results demonstrate the suitability of all three models for accurately predicting the energy demand of SW decoders.

Table~\ref{tab:SoftwareDecoderModeling}b presents the GPR evaluation. Once again, all energy models achieve an estimation error below \pro{2}, confirming their effectiveness for modeling SW decoders. The Valgrind 13PE model exhibits the lowest average error (\pro{1.38}), followed by the Temporal model (pro{1.73}). The Perf CTC model has the highest average error (\pro{1.89}) among the evaluated models.

The results suggest that all three models are effective for estimating SW decoder energy demand, with GPR providing a slight enhancement in accuracy compared to LR. These results also serve as a lower bound for the prediction accuracy in the subsequent HW decoder modeling and cross-codec prediction tasks.

\begin{table}[!t]

    \def\arraystretch{1.2}
	\centering
	\caption{MAPE results of the HW decoder energy demand modeling. The lowest $\overline{\varepsilon}$ for each decoder is given in bold.}
    \vspace{-0.4cm}
    \subfloat[Linear Regression]{
			
	\begin{tabular}{c l | r  | r  |  r }
		\multicolumn{2}{c|}{}  & \multicolumn{1}{c|}{}	& \multicolumn{1}{c|}{~~Perf~~}	& \multicolumn{1}{c}{Valgrind}	 \\
		\multicolumn{2}{c|}{Software Decoder}  & \multicolumn{1}{c|}{Temporal}	& \multicolumn{1}{c|}{ ~~CTC~~}	& \multicolumn{1}{c}{13PE} \\
	\hline \hline
    \multirow{2}{*}{AVC} & JM  & \pro{15.83} & \pro{23.56} & \textbf{\pro{3.70}} \\
                         & FFmpeg  & \pro{20.04} & \pro{21.02} & \textbf{\pro{3.77}} \\
    \hline
    \multirow{2}{*}{HEVC} & HM  & \pro{21.03} & \pro{25.01}  & \textbf{\pro{2.74}} \\
                         & openHEVC  & \pro{24.66} & \pro{28.53} & \textbf{\pro{4.02}} \\ 
    \hline
    \multirow{2}{*}{VP9} & libvpx  & \pro{21.10} & \pro{25.79}  & \textbf{\pro{5.50}} \\
                         & FFmpeg  & \pro{21.86} & \pro{38.97} & \textbf{\pro{3.65}} \\
    \hline
    \multirow{2}{*}{AV1} & libaom  & \pro{16.86} & \pro{23.72} &  \textbf{\pro{7.68}} \\ 
                         & dav1d  & \pro{16.10} & \pro{30.08} &  \textbf{\pro{6.66}} \\ 
    \hline
    \hline
     \multicolumn{2}{c|}{Average} & \pro{19.69} & \pro{27.09} &  \textbf{\pro{4.71}} \\
	\end{tabular}
	\label{tab:HardwareDecoderModelingLR}
    }
    \hspace{1.2cm}
    \subfloat[Gaussian Process Regression]{
        \begin{tabular}{c l | r  | r  |  r }
            \multicolumn{2}{c|}{}  & \multicolumn{1}{c|}{}	& \multicolumn{1}{c|}{~~Perf~~}	& \multicolumn{1}{c}{Valgrind}	 \\
            \multicolumn{2}{c|}{Software Decoder}  & \multicolumn{1}{c|}{Temporal}	& \multicolumn{1}{c|}{ ~~CTC~~}	& \multicolumn{1}{c}{13PE} \\
	\hline \hline
    \multirow{2}{*}{AVC} &  JM  & \pro{12.16} & \pro{5.99} & \textbf{\pro{2.53}} \\
                         & FFmpeg  & \pro{19.82} & \pro{7.06} &\textbf{\pro{2.52}} \\
    \hline
    \multirow{2}{*}{HEVC} & HM  & \pro{14.31} & \pro{7.54} &  \textbf{\pro{0.98}} \\
                         & openHEVC  & \pro{22.32} & \pro{9.24} &  \textbf{\pro{0.86}} \\
    \hline
    \multirow{2}{*}{VP9} & libvpx  & \pro{19.76} & \pro{10.81} & \textbf{\pro{0.93}} \\
                         & FFmpeg  & \pro{20.10} & \pro{8.06} &  \textbf{\pro{1.25}} \\ 
    \hline
    \multirow{2}{*}{AV1} & libaom  & \pro{20.59} & \pro{18.02} &  \textbf{\pro{2.82}} \\
                         & dav1d  & \pro{19.35} & \pro{15.67} & \textbf{\pro{2.46}} \\
    \hline
    \hline
  \multicolumn{2}{c|}{Average}  & \pro{18.55} & \pro{10.30} &  \textbf{\pro{~1.79}} \\
	\end{tabular}
	\label{tab:HardwareDecoderModelingGPR}
}
\label{tab:HardwareDecoderModeling}
\vspace{-0.4cm}
\end{table}

\section{Hardware Energy Modeling}
\label{sec:HardwareModeling}
\noindent 
In the following, we discuss the results for the HW decoder energy demand modeling, as outlined in the training phase of Fig.~\ref{fig:SolutionOverview}. Therefore, we take the SW decoders' profiling and decoding time measurement to estimate the HW decoder's energy demand. 

Table~\ref{tab:HardwareDecoderModeling}a summarizes the MAPE ($\overline{\varepsilon}$) for each model using LR. The results reveal that HW decoder modeling yields significantly higher $\overline{\varepsilon}$ compared to SW decoder modeling (c.f., Table~\ref{tab:SoftwareDecoderModeling}a). Specifically, the Temporal model exhibits an average $\overline{\varepsilon}$ of \pro{19.69}, a significant increase from \pro{1.57}. The Perf CTC model has an average $\overline{\varepsilon}$ of \pro{27.09} (c.f., \pro{2.75} in Table~\ref{tab:SoftwareDecoderModeling}a). These results indicate that neither model provides sufficient accuracy when using LR for HW decoder energy demand estimation. However, the Valgrind 13PE model achieves a comparatively lower $\overline{\varepsilon}$ of \pro{4.71}.

Switching to GPR significantly improves estimation accuracy, as demonstrated in Table~\ref{tab:HardwareDecoderModeling}b. Despite this improvement, the Temporal model still achieves a relatively high $\overline{\varepsilon}$ of \pro{18.55}, confirming that SW decoding time alone is insufficient for predicting HW decoder energy demand. The Perf CTC model also demonstrates high errors, with an average $\overline{\varepsilon}$ of \pro{10.30}. However, the Valgrind 13PE model reduces the error substantially to \pro{1.79}, comparable to the error levels observed for SW decoder modeling in Table~\ref{tab:SoftwareDecoderModeling}b. This demonstrates that using GPR instead of LR can reduce the estimation error by more than half, establishing GPR as the preferred regression approach for subsequent analyses.

The superior accuracy of the Valgrind 13PE model can be attributed to its inclusion of additional features, such as detailed cache-level metrics (e.g., L1 and LL cache) and branch prediction events, which are absent in the Temporal and Perf CTC models. These features allow Valgrind to simulate a more comprehensive representation of the CPU behavior, resulting in significantly more accurate energy demand estimates for HW decoders.

\section{Hardware Energy Cross-Codec Prediction For Unknown Hardware Decoders}
\label{sec:HardwarePrediction}
\noindent 
Finally, we propose a methodology to predict the energy demand of an unknown HW decoder implementation, as illustrated in the middle box of Fig.~\ref{fig:SolutionOverview}. This approach utilizes energy models trained on a video coding standard with existing HW and SW implementations. The verification process estimates the energy consumption of a different standard’s HW decoder while relying solely on its SW decoder for prediction. In this study, the objective is to demonstrate that the energy demand of the AV1 HW decoder can be cross-codec predicted using models trained on AVC, HEVC, and VP9.

\subsection{Linear Transformation for Verification}
\noindent
According to Fig.~\ref{fig:SolutionOverview}, the energy models are first trained using codec A. Subsequently, the SW profiling of codec B is utilized to estimate the cross-codec energy prediction $\hat{E}_{\textrm{cross}}$. These predictions are derived using the energy models trained on codec A, applying either LR or GPR.

To systematically evaluate the energy models, one video coding standard is reserved for verification, while the remaining standards are used for training. Table~\ref{tab:Phases} outlines all decoders and combinations of video decoders used for training. Each training configuration is referred to as a phase.

In phases 1–3, a single video decoder is employed to train the GPR coefficients, as detailed in the second column of Table~\ref{tab:Phases}. In phases 4–7, data sets from multiple video decoders are combined for training. The third column specifies the decoders verified in each phase.

\begin{table}[t!]
    \def\arraystretch{1.2}
    \caption{Overview of the cross-codec prediction phases. The video coding standards used for the GPR training are denoted in the second column and for verification in the third column.}

    \vspace*{-0.3cm}
    \label{tab:Phases}
    \begin{center}
    \begin{tabular}{  c | c | c } 
          Phase &  Training (Codec A) & Verification (Codec B) \\
         \hline\hline
         Phase 1  & AVC & AV1 \\
        \hdashline
        Phase 2  & HEVC & AV1 \\
        \hdashline
        Phase 3  & VP9 & AV1 \\
        \hline
        Phase 4  & AVC + HEVC & AV1 \\
        \hdashline
        Phase 5  & AVC + VP9 & AV1 \\
        \hdashline
        Phase 6  & AVC + HEVC + VP9 & AV1 \\
        \hdashline
        Phase 7  & HEVC + VP9 & AV1 \\
    \end{tabular}
    \end{center}
    \vspace*{-0.8cm}
\end{table}

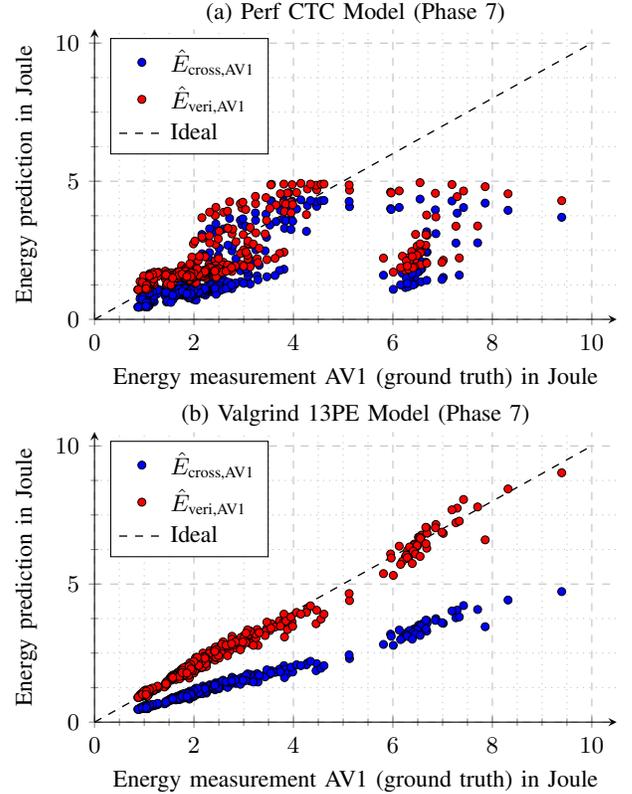
\begin{figure}[!t]

    \definecolor{SIMDoff}{HTML}{00BD47}
    \begin{center}
    \begin{tikzpicture}

    \tikzstyle{every node}=[font=\small]
    \begin{groupplot}[
         group style={group name=my plots,group size= 1 by 2,vertical sep =1.5cm},
          title style={yshift=-0.25cm},
         height = 10cm,
         width = \textwidth ,set layers,cell picture=true
        ]

        \nextgroupplot[
            width=0.47\textwidth,
            height = 0.3\textwidth,
               ylabel={Energy prediction in Joule},
               xlabel={Energy measurement AV1 (ground truth) in Joule},
           legend cell align = {left},
               title = {(a) Perf CTC Model (Phase 7)},
               xmin=0, xmax=10.5,
               ymin=0, ymax=10.5,
               axis lines = left,
               legend pos=north west,
               ymajorgrids=true,
               yminorgrids=true,
               xmajorgrids=true,
               xminorgrids=true,
               minor tick num=3,
               minor grid style=dotted,
               grid style = dashed,
                   yshift=0.33cm,
            ]
            
            \addplot[only marks,
            color=black,
            fill=blue,
             mark=*,
             mark size=1.5pt,
             line width=0.25pt,
             ]
             coordinates {
               (2.4181,1.1067)  (2.1565,1.0013)  (1.9409,1.0988)  (1.7855,0.95212)  (1.6651,0.94703)  (1.5253,0.89386)  (2.5064,1.2497)  (2.2606,1.128)  (2.0983,1.0739)  (2.0061,1.0399)  (1.9475,0.9266)  (1.8409,0.86159)  (2.4352,2.0997)  (2.1799,1.1432)  (1.9793,1.0127)  (1.8886,1.0306)  (1.7944,0.97635)  (1.6857,0.97765)  (2.3413,1.0623)  (2.2033,1.0689)  (2.1205,0.96191)  (2.0328,0.97327)  (1.9907,0.72173)  (1.9286,1.0131)  (2.4148,1.3861)  (2.2124,1.1076)  (2.0832,1.5036)  (1.9571,1.6347)  (1.8458,1.0463)  (1.7332,1.014)  (2.51,1.074)  (2.3559,1.182)  (2.1595,3.0732)  (2.0361,1.0305)  (2.0077,1.0409)  (1.9167,0.87803)  (2.9964,1.8946)  (2.6009,1.1934)  (2.336,1.0035)  (2.1125,1.0625)  (1.9418,0.93322)  (1.6515,0.90907)  (2.7556,1.2517)  (2.4757,1.4301)  (2.3033,1.1308)  (2.2323,0.87915)  (2.1643,0.93214)  (2.0679,0.94425)  (7.2925,3.8433)  (6.5816,1.8369)  (6.4268,1.4757)  (6.2958,1.6041)  (6.2511,1.1508)  (6.1906,1.6676)  (1.2042,0.96875)  (1.0656,0.78081)  (1.0091,0.91484)  (0.94763,0.63222)  (0.91636,0.74473)  (0.90242,0.46486)  (1.1874,0.97039)  (1.1169,1.0317)  (1.1072,0.66873)  (1.0449,0.65965)  (1.0595,0.51643)  (1.0614,0.45087)  (2.9372,1.7622)  (2.4234,1.022)  (2.0592,1.0781)  (1.7929,1.0174)  (1.6357,0.84517)  (1.4263,0.83499)  (2.8434,2.1935)  (2.4244,1.4899)  (2.1395,1.0374)  (1.9841,1.0042)  (1.8503,0.96757)  (1.9067,0.90848)  (6.6786,3.17)  (6.5142,1.7193)  (6.4366,1.5058)  (6.3783,1.8398)  (6.3411,1.3637)  (6.312,1.1858)  (4.2826,4.1219)  (3.65,3.5373)  (3.084,2.1374)  (2.5074,2.5762)  (2.3107,1.2534)  (1.9312,1.0307)  (3.9526,3.6359)  (3.2386,3.5854)  (2.8237,1.2671)  (2.4557,1.1294)  (2.2751,1.0194)  (2.0577,0.97344)  (3.8094,1.813)  (2.5904,1.4458)  (1.9565,1.0054)  (1.5806,0.72457)  (1.5419,0.86548)  (1.4493,0.63002)  (3.0858,2.2261)  (2.2354,1.0004)  (1.9183,0.89898)  (1.8504,0.89218)  (1.8707,0.89417)  (1.8592,0.86405)  (1.4238,1.1212)  (1.224,1.001)  (1.1444,0.80287)  (1.0706,0.91461)  (1.0416,0.94033)  (0.95897,0.55594)  (1.4037,0.96212)  (1.2628,1.0031)  (1.1968,0.94086)  (1.1419,0.60701)  (1.1172,0.81189)  (1.0937,0.48395)  (5.1282,4.079)  (4.2621,3.1864)  (3.271,1.5724)  (2.5748,1.3605)  (2.1786,1.059)  (1.7834,1.1102)  (4.4683,4.0928)  (3.7192,1.7666)  (3.0992,1.6059)  (2.4815,0.99583)  (2.1683,0.93023)  (1.9895,0.91647)  (7.1877,4.2842)  (6.864,3.1021)  (6.6859,2.4319)  (6.5657,1.3563)  (6.4867,2.0353)  (6.4362,1.2736)  (2.764,2.4818)  (2.3152,2.1548)  (2.0367,1.1126)  (1.8711,1.0983)  (1.779,1.1156)  (1.7053,0.91996)  (2.4937,1.5776)  (2.1674,1.0444)  (2.0855,1.0002)  (2.0118,0.97106)  (1.9562,0.95846)  (1.893,0.77531)  (2.0247,1.0799)  (1.8942,1.1285)  (1.763,1.0274)  (1.6263,0.97643)  (1.582,0.93815)  (1.5097,0.9078)  (2.2494,1.1182)  (2.1678,1.0921)  (2.042,0.69667)  (1.9431,1.0039)  (1.8672,0.94376)  (1.7724,0.87329)  (1.1483,0.98356)  (1.0218,0.76553)  (0.96807,0.49087)  (0.93673,0.46415)  (0.90121,0.45238)  (0.89402,0.44469)  (1.1086,0.56422)  (1.0305,0.53077)  (1.0175,0.49044)  (1.0484,0.47134)  (1.0292,0.4509)  (1.0392,0.45076)  (1.0939,0.98907)  (0.98737,0.76082)  (0.92704,0.49369)  (0.91706,0.47282)  (0.89658,0.45453)  (0.87051,0.44656)  (1.116,0.66906)  (1.0511,0.51936)  (1.0431,0.51245)  (1.0491,0.472)  (1.0548,0.47338)  (1.0551,0.449)  (2.6354,1.1946)  (2.3017,1.3561)  (2.0337,1.1288)  (1.8466,1.1309)  (1.7123,0.97467)  (1.561,0.92912)  (2.7092,1.2318)  (2.4574,1.1389)  (2.2515,1.1481)  (2.113,1.0443)  (2.0844,1.0001)  (2.0245,0.9249)  (2.2582,1.0634)  (1.9139,1.0286)  (1.6769,0.91411)  (1.5679,0.9222)  (1.5149,0.80298)  (1.4648,0.66929)  (2.0625,0.9903)  (1.9939,0.94388)  (1.9168,0.90298)  (1.8857,0.89226)  (1.8991,0.86962)  (1.9225,0.77587)  (3.5577,4.2919)  (3.5263,3.4142)  (3.2445,2.5187)  (2.7807,2.6609)  (2.4395,1.3252)  (1.9492,1.0842)  (3.909,3.9801)  (3.4593,2.9711)  (2.9912,2.4566)  (2.7283,1.3242)  (2.352,1.0932)  (2.0103,1.0449)  (3.8066,4.3002)  (3.2364,3.852)  (3.1305,2.7191)  (2.5944,3.2712)  (2.1749,2.3238)  (1.8005,1.0973)  (3.9736,4.2784)  (3.4482,3.808)  (3.0448,1.7855)  (2.6819,2.6854)  (2.2152,1.6801)  (1.846,1.0464)  (7.8558,4.2074)  (6.5463,4.3532)  (6.1273,4.052)  (5.9535,4.0085)  (5.9658,3.9767)  (5.8102,1.5979)  (9.3945,3.6953)  (8.3153,3.9471)  (7.7041,2.7696)  (7.2678,2.7623)  (7.0128,1.5174)  (6.6807,1.4115)  (5.1203,4.2689)  (2.9661,3.4283)  (2.2553,1.0856)  (1.9555,1.0646)  (1.8194,1.0175)  (1.5371,0.61412)  (3.7985,3.6163)  (2.2449,1.0557)  (2.0063,0.95046)  (1.9725,0.87603)  (1.9313,0.96673)  (1.9081,0.80362)  (2.3686,1.137)  (1.9897,1.1072)  (1.8239,1.0399)  (1.7409,0.9827)  (1.6341,0.94087)  (1.5524,0.91531)  (2.1661,1.0629)  (2.0367,1.1071)  (1.92,0.97497)  (1.8538,0.94102)  (1.7786,0.90122)  (1.8275,0.86378)  (6.8545,3.9647)  (6.5374,1.859)  (6.3357,1.5124)  (6.2701,1.4894)  (6.1547,1.2491)  (6.0083,1.0844)  (2.0628,1.0723)  (1.985,1.0756)  (1.8813,0.9022)  (1.8141,0.88915)  (1.704,1.0001)  (1.6363,0.86817)  (2.1472,1.1063)  (2.0426,1.0507)  (1.9907,1.0158)  (1.9231,1.004)  (1.9441,0.99538)  (1.9066,0.94279)  (2.9338,3.5888)  (2.4269,3.1555)  (1.9885,2.1192)  (1.7521,1.319)  (1.6756,1.1033)  (1.5729,0.98667)  (2.7161,3.6566)  (2.3144,3.2771)  (2.0196,2.218)  (1.8501,1.2398)  (1.7531,1.1119)  (1.7457,1.1137)  (1.198,1.0818)  (1.1125,0.93394)  (1.071,0.96355)  (1.0146,0.82337)  (1.0049,0.76934)  (0.97064,0.52808)  (1.2643,0.92531)  (1.1995,0.63494)  (1.1288,0.89376)  (1.0904,0.82818)  (1.0822,0.86437)  (1.049,0.47511)  (3.8463,3.5731)  (3.0861,1.562)  (2.4294,1.1993)  (2.0405,1.0891)  (1.8486,1.0397)  (1.6173,0.97225)  (3.6295,1.5887)  (2.7613,1.1119)  (2.3163,1.2741)  (2.1805,1.1298)  (2.0746,1.0637)  (1.9955,0.95333)  (1.8388,1.0883)  (1.7928,1.0477)  (1.7692,0.9655)  (1.7371,1.0932)  (1.6377,0.95671)  (1.5312,1.0216)  (1.8775,1.0101)  (1.8665,1.0056)  (1.8534,1.0184)  (1.8121,0.9461)  (1.793,0.91789)  (1.7134,0.89681)  (4.4557,4.2572)  (3.8794,3.9744)  (3.8207,3.2977)  (3.2872,1.4309)  (2.6752,1.3468)  (1.9974,0.99217)  (4.5551,4.2936)  (3.9849,3.5506)  (3.4304,2.4796)  (3.1944,2.111)  (2.505,1.2117)  (1.9544,1.0327)  (2.2223,1.6455)  (1.848,1.2263)  (1.6535,1.0773)  (1.5437,1.0618)  (1.5634,0.97755)  (1.5821,0.94006)  (1.9835,1.2423)  (1.7493,1.1038)  (1.6734,1.0827)  (1.6591,1.0493)  (1.7007,1.0932)  (1.7615,0.96744)  (3.1293,1.2817)  (2.9421,1.3637)  (2.5486,1.0685)  (2.3445,1.2359)  (2.1403,1.0187)  (1.7352,0.94554)  (3.312,1.9588)  (2.9997,1.3043)  (2.7033,1.1692)  (2.4445,0.908)  (2.1919,1.0482)  (2.0021,0.92944)  (7.4207,4.0484)  (6.6621,2.069)  (6.5533,1.7973)  (6.4844,1.6986)  (6.4051,1.2465)  (6.2893,1.3061)  (7.3351,1.6012)  (6.9862,1.4431)  (6.6652,2.4501)  (6.4557,2.4636)  (6.5471,2.1182)  (6.3888,1.6524)  (4.3429,4.2999)  (3.9516,3.2521)  (3.0775,1.7053)  (2.5846,1.1005)  (2.1799,1.0797)  (1.8217,0.99374)  (4.0636,3.9932)  (3.4669,1.5414)  (3.0478,1.8181)  (2.563,1.0822)  (2.208,0.94339)  (2.0183,0.91107)  (4.1489,4.3259)  (3.5465,4.2698)  (3.0411,3.3718)  (2.4402,3.4514)  (2.1604,2.3747)  (1.8899,1.2543)  (4.614,4.3029)  (3.7383,3.9509)  (2.9828,3.6387)  (2.4884,2.3007)  (2.2385,1.1649)  (2.0112,1.0847)  (2.6736,3.313)  (2.3117,2.4149)  (2.1112,1.6643)  (2.0281,1.1154)  (1.9364,1.1367)  (1.8251,1.1116)  (2.6453,1.1199)  (2.4046,1.129)  (2.2338,1.057)  (2.1626,0.95692)  (2.0825,0.95041)  (1.9451,0.95233)  (1.1422,0.65571)  (1.0122,0.69716)  (0.98306,0.53262)  (0.93098,0.50807)  (0.91866,0.46156)  (0.89199,0.44333)  (1.1835,0.69403)  (1.1067,0.5361)  (1.0649,0.51689)  (1.0578,0.48874)  (1.0642,0.47008)  (1.049,0.44747)  (1.1419,0.71676)  (1.0461,0.66666)  (0.97579,0.53049)  (0.93946,0.48326)  (0.90893,0.45082)  (0.88007,0.44537)  (1.1902,0.84182)  (1.084,0.53281)  (1.0603,0.51888)  (1.0581,0.51213)  (1.0474,0.46537)  (1.0621,0.45068)  (1.4561,1.0368)  (1.2872,1.0832)  (1.1442,0.9807)  (1.0275,0.93868)  (0.98921,0.94856)  (0.96258,0.71751)  (1.4373,0.98461)  (1.2663,1.0547)  (1.1659,0.99541)  (1.1048,0.9449)  (1.045,0.60483)  (1.0375,0.48555)  (2.8339,1.314)  (2.4441,1.1733)  (2.2041,1.0533)  (2.0881,1.051)  (1.8789,0.99695)  (1.7037,0.96633)  (2.7565,1.2511)  (2.4196,1.1213)  (2.2203,1.1056)  (2.0871,0.99792)  (2.097,0.98544)  (1.9667,0.89535)  (2.0266,1.1208)  (1.9164,1.3881)  (1.7929,0.98018)  (1.7689,0.99489)  (1.7626,1.0126)  (1.573,0.87303)  (2.0657,1.2084)  (1.9966,1.0432)  (1.9377,1.0528)  (1.9011,0.981)  (1.9401,0.92633)  (1.8569,0.92728)  
             };
             \addlegendentry{$\hat{E}_{\textrm{cross,AV1}}$}

         \addplot[only marks,
            color=black,
            fill=red,
             mark=*,
             mark size=1.5pt,
             line width=0.25pt,
             ]
             coordinates {
                
        (2.4181,1.7309)  (2.1565,1.6264)  (1.9409,1.723)  (1.7855,1.5776)  (1.6651,1.5726)  (1.5253,1.5199)  (2.5064,1.8725)  (2.2606,1.752)  (2.0983,1.6984)  (2.0061,1.6647)  (1.9475,1.5524)  (1.8409,1.4879)  (2.4352,2.715)  (2.1799,1.767)  (1.9793,1.6377)  (1.8886,1.6554)  (1.7944,1.6017)  (1.6857,1.6029)  (2.3413,1.6868)  (2.2033,1.6934)  (2.1205,1.5873)  (2.0328,1.5986)  (1.9907,1.3493)  (1.9286,1.6381)  (2.4148,2.0077)  (2.2124,1.7317)  (2.0832,2.1242)  (1.9571,2.2541)  (1.8458,1.671)  (1.7332,1.639)  (2.51,1.6984)  (2.3559,1.8055)  (2.1595,3.6798)  (2.0361,1.6553)  (2.0077,1.6657)  (1.9167,1.5042)  (2.9964,2.5117)  (2.6009,1.8167)  (2.336,1.6286)  (2.1125,1.687)  (1.9418,1.5589)  (1.6515,1.535)  (2.7556,1.8746)  (2.4757,2.0514)  (2.3033,1.7547)  (2.2323,1.5053)  (2.1643,1.5578)  (2.0679,1.5698)  (7.2925,4.443)  (6.5816,2.4545)  (6.4268,2.0966)  (6.2958,2.2238)  (6.2511,1.7746)  (6.1906,2.2868)  (1.2042,1.5941)  (1.0656,1.4079)  (1.0091,1.5407)  (0.94763,1.2606)  (0.91636,1.3721)  (0.90242,1.0947)  (1.1874,1.5957)  (1.1169,1.6565)  (1.1072,1.2968)  (1.0449,1.2878)  (1.0595,1.1458)  (1.0614,1.0809)  (2.9372,2.3805)  (2.4234,1.6469)  (2.0592,1.7025)  (1.7929,1.6423)  (1.6357,1.4716)  (1.4263,1.4616)  (2.8434,2.808)  (2.4244,2.1107)  (2.1395,1.6622)  (1.9841,1.6292)  (1.8503,1.593)  (1.9067,1.5344)  (6.6786,3.7758)  (6.5142,2.338)  (6.4366,2.1264)  (6.3783,2.4574)  (6.3411,1.9856)  (6.312,1.8092)  (4.2826,4.7191)  (3.65,4.1398)  (3.084,2.7523)  (2.5074,3.1873)  (2.3107,1.8763)  (1.9312,1.6555)  (3.9526,4.2375)  (3.2386,4.1875)  (2.8237,1.8899)  (2.4557,1.7533)  (2.2751,1.6443)  (2.0577,1.5988)  (3.8094,2.4308)  (2.5904,2.0669)  (1.9565,1.6304)  (1.5806,1.3521)  (1.5419,1.4918)  (1.4493,1.2584)  (3.0858,2.8403)  (2.2354,1.6255)  (1.9183,1.525)  (1.8504,1.5182)  (1.8707,1.5202)  (1.8592,1.4904)  (1.4238,1.7453)  (1.224,1.6261)  (1.1444,1.4297)  (1.0706,1.5405)  (1.0416,1.566)  (0.95897,1.185)  (1.4037,1.5876)  (1.2628,1.6282)  (1.1968,1.5665)  (1.1419,1.2356)  (1.1172,1.4387)  (1.0937,1.1136)  (5.1282,4.6767)  (4.2621,3.792)  (3.271,2.1924)  (2.5748,1.9824)  (2.1786,1.6836)  (1.7834,1.7343)  (4.4683,4.6904)  (3.7192,2.3849)  (3.0992,2.2256)  (2.4815,1.621)  (2.1683,1.5559)  (1.9895,1.5423)  (7.1877,4.88)  (6.864,3.7085)  (6.6859,3.0443)  (6.5657,1.9782)  (6.4867,2.6512)  (6.4362,1.8963)  (2.764,3.0937)  (2.3152,2.7696)  (2.0367,1.7367)  (1.8711,1.7226)  (1.779,1.7396)  (1.7053,1.5458)  (2.4937,2.1976)  (2.1674,1.6692)  (2.0855,1.6253)  (2.0118,1.5964)  (1.9562,1.5839)  (1.893,1.4024)  (2.0247,1.7043)  (1.8942,1.7525)  (1.763,1.6523)  (1.6263,1.6017)  (1.582,1.5638)  (1.5097,1.5337)  (2.2494,1.7422)  (2.1678,1.7164)  (2.042,1.3245)  (1.9431,1.629)  (1.8672,1.5694)  (1.7724,1.4995)  (1.1483,1.6088)  (1.0218,1.3927)  (0.96807,1.1205)  (0.93673,1.094)  (0.90121,1.0824)  (0.89402,1.0747)  (1.1086,1.1932)  (1.0305,1.16)  (1.0175,1.1201)  (1.0484,1.1011)  (1.0292,1.0809)  (1.0392,1.0808)  (1.0939,1.6143)  (0.98737,1.3881)  (0.92704,1.1233)  (0.91706,1.1026)  (0.89658,1.0845)  (0.87051,1.0766)  (1.116,1.2971)  (1.0511,1.1487)  (1.0431,1.1419)  (1.0491,1.1018)  (1.0548,1.1032)  (1.0551,1.079)  (2.6354,1.8179)  (2.3017,1.978)  (2.0337,1.7527)  (1.8466,1.7549)  (1.7123,1.6)  (1.561,1.5549)  (2.7092,1.8549)  (2.4574,1.7628)  (2.2515,1.7719)  (2.113,1.669)  (2.0844,1.6251)  (2.0245,1.5507)  (2.2582,1.688)  (1.9139,1.6535)  (1.6769,1.54)  (1.5679,1.548)  (1.5149,1.4298)  (1.4648,1.2973)  (2.0625,1.6155)  (1.9939,1.5695)  (1.9168,1.5289)  (1.8857,1.5183)  (1.8991,1.4959)  (1.9225,1.403)  (3.5577,4.8876)  (3.5263,4.0177)  (3.2445,3.1303)  (2.7807,3.2712)  (2.4395,1.9474)  (1.9492,1.7086)  (3.909,4.5787)  (3.4593,3.5786)  (2.9912,3.0688)  (2.7283,1.9464)  (2.352,1.7174)  (2.0103,1.6696)  (3.8066,4.8959)  (3.2364,4.4516)  (3.1305,3.3289)  (2.5944,3.876)  (2.1749,2.9371)  (1.8005,1.7215)  (3.9736,4.8743)  (3.4482,4.4081)  (3.0448,2.4036)  (2.6819,3.2955)  (2.2152,2.2991)  (1.846,1.671)  (7.8558,4.8039)  (6.5463,4.9484)  (6.1273,4.6499)  (5.9535,4.6068)  (5.9658,4.5753)  (5.8102,2.2177)  (9.3945,4.2964)  (8.3153,4.5459)  (7.7041,3.3789)  (7.2678,3.3717)  (7.0128,2.1379)  (6.6807,2.033)  (5.1203,4.8649)  (2.9661,4.0317)  (2.2553,1.7099)  (1.9555,1.6891)  (1.8194,1.6424)  (1.5371,1.2427)  (3.7985,4.218)  (2.2449,1.6803)  (2.0063,1.576)  (1.9725,1.5022)  (1.9313,1.5921)  (1.9081,1.4305)  (2.3686,1.7609)  (1.9897,1.7313)  (1.8239,1.6646)  (1.7409,1.6079)  (1.6341,1.5665)  (1.5524,1.5412)  (2.1661,1.6874)  (2.0367,1.7313)  (1.92,1.6003)  (1.8538,1.5666)  (1.7786,1.5272)  (1.8275,1.4901)  (6.8545,4.5634)  (6.5374,2.4764)  (6.3357,2.1329)  (6.2701,2.1101)  (6.1547,1.872)  (6.0083,1.7087)  (2.0628,1.6968)  (1.985,1.7)  (1.8813,1.5282)  (1.8141,1.5152)  (1.704,1.6252)  (1.6363,1.4944)  (2.1472,1.7304)  (2.0426,1.6753)  (1.9907,1.6408)  (1.9231,1.6291)  (1.9441,1.6205)  (1.9066,1.5684)  (2.9338,4.1909)  (2.4269,3.7614)  (1.9885,2.7343)  (1.7521,1.9413)  (1.6756,1.7275)  (1.5729,1.6119)  (2.7161,4.258)  (2.3144,3.8819)  (2.0196,2.8322)  (1.8501,1.8627)  (1.7531,1.736)  (1.7457,1.7378)  (1.198,1.7061)  (1.1125,1.5596)  (1.071,1.589)  (1.0146,1.45)  (1.0049,1.3965)  (0.97064,1.1574)  (1.2643,1.5511)  (1.1995,1.2633)  (1.1288,1.5198)  (1.0904,1.4548)  (1.0822,1.4907)  (1.049,1.1049)  (3.8463,4.1752)  (3.0861,2.1821)  (2.4294,1.8226)  (2.0405,1.7134)  (1.8486,1.6644)  (1.6173,1.5976)  (3.6295,2.2086)  (2.7613,1.736)  (2.3163,1.8968)  (2.1805,1.7538)  (2.0746,1.6882)  (1.9955,1.5788)  (1.8388,1.7126)  (1.7928,1.6723)  (1.7692,1.5909)  (1.7371,1.7174)  (1.6377,1.5822)  (1.5312,1.6465)  (1.8775,1.6351)  (1.8665,1.6307)  (1.8534,1.6433)  (1.8121,1.5717)  (1.793,1.5437)  (1.7134,1.5228)  (4.4557,4.8533)  (3.8794,4.573)  (3.8207,3.9023)  (3.2872,2.0521)  (2.6752,1.9688)  (1.9974,1.6173)  (4.5551,4.8893)  (3.9849,4.153)  (3.4304,3.0915)  (3.1944,2.7262)  (2.505,1.8349)  (1.9544,1.6575)  (2.2223,2.2648)  (1.848,1.8494)  (1.6535,1.7017)  (1.5437,1.6863)  (1.5634,1.6029)  (1.5821,1.5657)  (1.9835,1.8652)  (1.7493,1.7279)  (1.6734,1.707)  (1.6591,1.674)  (1.7007,1.7175)  (1.7615,1.5928)  (3.1293,1.9043)  (2.9421,1.9856)  (2.5486,1.693)  (2.3445,1.8589)  (2.1403,1.6437)  (1.7352,1.5711)  (3.312,2.5754)  (2.9997,1.9267)  (2.7033,1.7928)  (2.4445,1.5339)  (2.1919,1.6729)  (2.0021,1.5552)  (7.4207,4.6464)  (6.6621,2.6845)  (6.5533,2.4153)  (6.4844,2.3175)  (6.4051,1.8694)  (6.2893,1.9285)  (7.3351,2.2209)  (6.9862,2.0643)  (6.6652,3.0623)  (6.4557,3.0757)  (6.5471,2.7334)  (6.3888,2.2717)  (4.3429,4.8956)  (3.9516,3.8571)  (3.0775,2.3241)  (2.5846,1.7247)  (2.1799,1.7041)  (1.8217,1.6189)  (4.0636,4.5917)  (3.4669,2.1617)  (3.0478,2.4359)  (2.563,1.7065)  (2.208,1.569)  (2.0183,1.537)  (4.1489,4.9214)  (3.5465,4.8658)  (3.0411,3.9758)  (2.4402,4.0547)  (2.1604,2.9876)  (1.8899,1.8772)  (4.614,4.8985)  (3.7383,4.5497)  (2.9828,4.2402)  (2.4884,2.9142)  (2.2385,1.7886)  (2.0112,1.709)  (2.6736,3.9175)  (2.3117,3.0274)  (2.1112,2.2835)  (2.0281,1.7395)  (1.9364,1.7605)  (1.8251,1.7357)  (2.6453,1.744)  (2.4046,1.753)  (2.2338,1.6816)  (2.1626,1.5824)  (2.0825,1.576)  (1.9451,1.5779)  (1.1422,1.2839)  (1.0122,1.325)  (0.98306,1.1619)  (0.93098,1.1376)  (0.91866,1.0915)  (0.89199,1.0734)  (1.1835,1.3219)  (1.1067,1.1653)  (1.0649,1.1463)  (1.0578,1.1184)  (1.0642,1.0999)  (1.049,1.0775)  (1.1419,1.3444)  (1.0461,1.2947)  (0.97579,1.1598)  (0.93946,1.113)  (0.90893,1.0808)  (0.88007,1.0754)  (1.1902,1.4683)  (1.084,1.1621)  (1.0603,1.1483)  (1.0581,1.1416)  (1.0474,1.0952)  (1.0621,1.0807)  (1.4561,1.6615)  (1.2872,1.7075)  (1.1442,1.606)  (1.0275,1.5643)  (0.98921,1.5741)  (0.96258,1.3451)  (1.4373,1.6098)  (1.2663,1.6793)  (1.1659,1.6206)  (1.1048,1.5705)  (1.045,1.2334)  (1.0375,1.1152)  (2.8339,1.9363)  (2.4441,1.7969)  (2.2041,1.6779)  (2.0881,1.6757)  (1.8789,1.6221)  (1.7037,1.5917)  (2.7565,1.874)  (2.4196,1.7453)  (2.2203,1.7298)  (2.0871,1.623)  (2.097,1.6107)  (1.9667,1.5214)  (2.0266,1.7449)  (1.9164,2.0097)  (1.7929,1.6055)  (1.7689,1.62)  (1.7626,1.6376)  (1.573,1.4993)  (2.0657,1.8317)  (1.9966,1.6679)  (1.9377,1.6774)  (1.9011,1.6063)  (1.9401,1.5521)  (1.8569,1.553)  
             };
             \addlegendentry{$\hat{E}_{\textrm{veri,AV1}}$}

             \addplot[
             color=black,
              mark=none,
              line width=0.5pt,dashed
              ]
              coordinates {
            (0,0)(10,10)
              };
              \addlegendentry{Ideal}

        \nextgroupplot[
            width=0.47\textwidth,
            height = 0.3\textwidth,
               ylabel={Energy prediction in Joule},
               xlabel={Energy measurement AV1 (ground truth) in Joule},
           legend cell align = {left},
              title = {(b) Valgrind 13PE Model (Phase 7)},
               xmin=0, xmax=10.5,
               ymin=0, ymax=10.5,
               axis lines = left,
               legend pos=north west,
               ymajorgrids=true,
               yminorgrids=true,
               xmajorgrids=true,
               xminorgrids=true,
               minor tick num=3,
               minor grid style=dotted,
               grid style = dashed,
            ]
            
            \addplot[only marks,
            color=black,
            fill=blue,
             mark=*,
             mark size=1.5pt,
             line width=0.25pt,
             ]
             coordinates {
                (2.4181,1.2796)  (2.1565,1.1423)  (1.9409,1.0218)  (1.7855,0.9442)  (1.6651,0.88655)  (1.5253,0.81463)  (2.5064,1.2791)  (2.2606,1.1641)  (2.0983,1.0766)  (2.0061,1.0153)  (1.9475,0.97192)  (1.8409,0.91421)  (2.4352,1.2902)  (2.1799,1.1049)  (1.9793,1.0209)  (1.8886,0.9302)  (1.7944,0.87291)  (1.6857,0.82425)  (2.3413,1.2057)  (2.2033,1.1241)  (2.1205,1.0629)  (2.0328,1.0212)  (1.9907,0.98129)  (1.9286,0.91869)  (2.4148,1.3301)  (2.2124,1.2469)  (2.0832,1.1487)  (1.9571,1.057)  (1.8458,0.9743)  (1.7332,0.87896)  (2.51,1.3559)  (2.3559,1.2682)  (2.1595,1.1613)  (2.0361,1.0748)  (2.0077,1.0236)  (1.9167,0.92038)  (2.9964,1.6305)  (2.6009,1.4525)  (2.336,1.3233)  (2.1125,1.1503)  (1.9418,1.0258)  (1.6515,0.8748)  (2.7556,1.4381)  (2.4757,1.3143)  (2.3033,1.215)  (2.2323,1.1432)  (2.1643,1.09)  (2.0679,1.0095)  (7.2925,4.0552)  (6.5816,3.5461)  (6.4268,3.3729)  (6.2958,3.2562)  (6.2511,3.1336)  (6.1906,3.0452)  (1.2042,0.63053)  (1.0656,0.59371)  (1.0091,0.56617)  (0.94763,0.54139)  (0.91636,0.51604)  (0.90242,0.47577)  (1.1874,0.61716)  (1.1169,0.59273)  (1.1072,0.57177)  (1.0449,0.54938)  (1.0595,0.52648)  (1.0614,0.49135)  (2.9372,1.4938)  (2.4234,1.2071)  (2.0592,1.0235)  (1.7929,0.89714)  (1.6357,0.80493)  (1.4263,0.73274)  (2.8434,1.4021)  (2.4244,1.1719)  (2.1395,1.0332)  (1.9841,0.95308)  (1.8503,0.89075)  (1.9067,0.83262)  (6.6786,3.6827)  (6.5142,3.5039)  (6.4366,3.3577)  (6.3783,3.2619)  (6.3411,3.1943)  (6.312,3.1007)  (4.2826,2.1387)  (3.65,1.8171)  (3.084,1.6213)  (2.5074,1.4337)  (2.3107,1.2544)  (1.9312,0.98371)  (3.9526,1.9678)  (3.2386,1.7209)  (2.8237,1.5618)  (2.4557,1.4005)  (2.2751,1.2465)  (2.0577,1.0823)  (3.8094,2.0449)  (2.5904,1.3374)  (1.9565,0.9785)  (1.5806,0.85767)  (1.5419,0.82569)  (1.4493,0.76246)  (3.0858,1.6563)  (2.2354,1.1614)  (1.9183,0.97027)  (1.8504,0.93848)  (1.8707,0.93427)  (1.8592,0.90342)  (1.4238,0.6781)  (1.224,0.61744)  (1.1444,0.59889)  (1.0706,0.57216)  (1.0416,0.53885)  (0.95897,0.49042)  (1.4037,0.66317)  (1.2628,0.6292)  (1.1968,0.6064)  (1.1419,0.5862)  (1.1172,0.56333)  (1.0937,0.52997)  (5.1282,2.3022)  (4.2621,2.0765)  (3.271,1.7531)  (2.5748,1.3944)  (2.1786,1.1029)  (1.7834,0.88294)  (4.4683,2.119)  (3.7192,1.9179)  (3.0992,1.616)  (2.4815,1.2811)  (2.1683,1.0983)  (1.9895,0.97989)  (7.1877,4.0235)  (6.864,3.7448)  (6.6859,3.5826)  (6.5657,3.4477)  (6.4867,3.3335)  (6.4362,3.2396)  (2.764,1.6249)  (2.3152,1.2283)  (2.0367,1.0222)  (1.8711,0.9307)  (1.779,0.88359)  (1.7053,0.85262)  (2.4937,1.375)  (2.1674,1.1606)  (2.0855,1.1022)  (2.0118,1.0517)  (1.9562,1.0062)  (1.893,0.94696)  (2.0247,0.99271)  (1.8942,0.96573)  (1.763,0.94658)  (1.6263,0.90659)  (1.582,0.87269)  (1.5097,0.82102)  (2.2494,1.0516)  (2.1678,1.0335)  (2.042,1.0033)  (1.9431,0.97943)  (1.8672,0.94418)  (1.7724,0.88993)  (1.1483,0.59886)  (1.0218,0.54121)  (0.96807,0.50791)  (0.93673,0.48697)  (0.90121,0.47816)  (0.89402,0.46831)  (1.1086,0.57524)  (1.0305,0.53717)  (1.0175,0.52101)  (1.0484,0.50687)  (1.0292,0.49595)  (1.0392,0.48704)  (1.0939,0.58121)  (0.98737,0.53945)  (0.92704,0.50476)  (0.91706,0.48627)  (0.89658,0.47204)  (0.87051,0.46484)  (1.116,0.57774)  (1.0511,0.5483)  (1.0431,0.52651)  (1.0491,0.50735)  (1.0548,0.49565)  (1.0551,0.48773)  (2.6354,1.459)  (2.3017,1.3114)  (2.0337,1.1707)  (1.8466,1.0477)  (1.7123,0.96079)  (1.561,0.86535)  (2.7092,1.4258)  (2.4574,1.3059)  (2.2515,1.2039)  (2.113,1.1223)  (2.0844,1.085)  (2.0245,1.0133)  (2.2582,1.1385)  (1.9139,0.96668)  (1.6769,0.86967)  (1.5679,0.82501)  (1.5149,0.80293)  (1.4648,0.76725)  (2.0625,1.0507)  (1.9939,1.0029)  (1.9168,0.96504)  (1.8857,0.93086)  (1.8991,0.91707)  (1.9225,0.90513)  (3.5577,1.8637)  (3.5263,1.734)  (3.2445,1.5862)  (2.7807,1.4593)  (2.4395,1.3232)  (1.9492,1.0323)  (3.909,1.9054)  (3.4593,1.767)  (2.9912,1.6062)  (2.7283,1.4728)  (2.352,1.3154)  (2.0103,1.0657)  (3.8066,1.8374)  (3.2364,1.6676)  (3.1305,1.4921)  (2.5944,1.3418)  (2.1749,1.1993)  (1.8005,0.93819)  (3.9736,1.8951)  (3.4482,1.7253)  (3.0448,1.5452)  (2.6819,1.3888)  (2.2152,1.2244)  (1.846,0.97813)  (7.8558,3.454)  (6.5463,3.4007)  (6.1273,3.3341)  (5.9535,3.1859)  (5.9658,3.11)  (5.8102,2.8138)  (9.3945,4.7261)  (8.3153,4.4197)  (7.7041,4.078)  (7.2678,3.7784)  (7.0128,3.5772)  (6.6807,3.2964)  (5.1203,2.4352)  (2.9661,1.4428)  (2.2553,1.1198)  (1.9555,0.9713)  (1.8194,0.89036)  (1.5371,0.78933)  (3.7985,1.8886)  (2.2449,1.1575)  (2.0063,1.0238)  (1.9725,0.97411)  (1.9313,0.9459)  (1.9081,0.92492)  (2.3686,1.3027)  (1.9897,1.1432)  (1.8239,1.028)  (1.7409,0.95242)  (1.6341,0.88468)  (1.5524,0.82457)  (2.1661,1.1863)  (2.0367,1.0951)  (1.92,1.0284)  (1.8538,0.97195)  (1.7786,0.92883)  (1.8275,0.88385)  (6.8545,3.6831)  (6.5374,3.5066)  (6.3357,3.3177)  (6.2701,3.165)  (6.1547,2.9892)  (6.0083,2.7804)  (2.0628,1.1613)  (1.985,1.087)  (1.8813,1.0232)  (1.8141,0.9721)  (1.704,0.92053)  (1.6363,0.85719)  (2.1472,1.1653)  (2.0426,1.1079)  (1.9907,1.0603)  (1.9231,1.0181)  (1.9441,1.0019)  (1.9066,0.95789)  (2.9338,1.3691)  (2.4269,1.2277)  (1.9885,1.0948)  (1.7521,1.0045)  (1.6756,0.93943)  (1.5729,0.88662)  (2.7161,1.3437)  (2.3144,1.2317)  (2.0196,1.116)  (1.8501,1.0378)  (1.7531,0.97208)  (1.7457,0.92617)  (1.198,0.62645)  (1.1125,0.60083)  (1.071,0.57997)  (1.0146,0.56055)  (1.0049,0.53485)  (0.97064,0.48945)  (1.2643,0.65474)  (1.1995,0.62831)  (1.1288,0.6042)  (1.0904,0.58095)  (1.0822,0.55565)  (1.049,0.50524)  (3.8463,1.9143)  (3.0861,1.6548)  (2.4294,1.4194)  (2.0405,1.2236)  (1.8486,1.053)  (1.6173,0.87824)  (3.6295,1.802)  (2.7613,1.5708)  (2.3163,1.35)  (2.1805,1.2218)  (2.0746,1.1292)  (1.9955,1.014)  (1.8388,1.0202)  (1.7928,1.0031)  (1.7692,0.96711)  (1.7371,0.94095)  (1.6377,0.88999)  (1.5312,0.8197)  (1.8775,1.0136)  (1.8665,0.98939)  (1.8534,0.9678)  (1.8121,0.93511)  (1.793,0.89588)  (1.7134,0.81766)  (4.4557,1.863)  (3.8794,1.7516)  (3.8207,1.6118)  (3.2872,1.4529)  (2.6752,1.2581)  (1.9974,0.92784)  (4.5551,1.952)  (3.9849,1.8144)  (3.4304,1.6403)  (3.1944,1.4535)  (2.505,1.2126)  (1.9544,0.97252)  (2.2223,1.1456)  (1.848,0.99911)  (1.6535,0.90377)  (1.5437,0.86349)  (1.5634,0.85365)  (1.5821,0.82837)  (1.9835,1.0288)  (1.7493,0.94228)  (1.6734,0.89459)  (1.6591,0.88488)  (1.7007,0.88133)  (1.7615,0.87364)  (3.1293,1.7254)  (2.9421,1.5999)  (2.5486,1.4346)  (2.3445,1.2745)  (2.1403,1.1089)  (1.7352,0.89249)  (3.312,1.6951)  (2.9997,1.5854)  (2.7033,1.4351)  (2.4445,1.2737)  (2.1919,1.1322)  (2.0021,0.98863)  (7.4207,4.2193)  (6.6621,3.6913)  (6.5533,3.4823)  (6.4844,3.3446)  (6.4051,3.2404)  (6.2893,3.1187)  (7.3351,3.809)  (6.9862,3.5998)  (6.6652,3.3798)  (6.4557,3.1842)  (6.5471,3.14)  (6.3888,3.0061)  (4.3429,2.2024)  (3.9516,1.9907)  (3.0775,1.7463)  (2.5846,1.5193)  (2.1799,1.3063)  (1.8217,1.0296)  (4.0636,2.0768)  (3.4669,1.8931)  (3.0478,1.6844)  (2.563,1.4751)  (2.208,1.246)  (2.0183,1.0473)  (4.1489,2.0453)  (3.5465,1.778)  (3.0411,1.5076)  (2.4402,1.2291)  (2.1604,1.0374)  (1.8899,0.88287)  (4.614,2.0446)  (3.7383,1.8176)  (2.9828,1.5157)  (2.4884,1.2603)  (2.2385,1.1063)  (2.0112,0.97148)  (2.6736,1.4405)  (2.3117,1.3036)  (2.1112,1.1699)  (2.0281,1.0575)  (1.9364,0.97906)  (1.8251,0.89855)  (2.6453,1.4554)  (2.4046,1.3484)  (2.2338,1.2292)  (2.1626,1.1557)  (2.0825,1.099)  (1.9451,1.0055)  (1.1422,0.5885)  (1.0122,0.54371)  (0.98306,0.51843)  (0.93098,0.49803)  (0.91866,0.48084)  (0.89199,0.46753)  (1.1835,0.58929)  (1.1067,0.56151)  (1.0649,0.53428)  (1.0578,0.51502)  (1.0642,0.50099)  (1.049,0.49012)  (1.1419,0.60443)  (1.0461,0.56328)  (0.97579,0.52459)  (0.93946,0.49858)  (0.90893,0.48006)  (0.88007,0.46578)  (1.1902,0.6063)  (1.084,0.57603)  (1.0603,0.54413)  (1.0581,0.51912)  (1.0474,0.50445)  (1.0621,0.48904)  (1.4561,0.71227)  (1.2872,0.66563)  (1.1442,0.63205)  (1.0275,0.60614)  (0.98921,0.58047)  (0.96258,0.52461)  (1.4373,0.72362)  (1.2663,0.67918)  (1.1659,0.64234)  (1.1048,0.6158)  (1.045,0.5826)  (1.0375,0.52179)  (2.8339,1.519)  (2.4441,1.3711)  (2.2041,1.2339)  (2.0881,1.0972)  (1.8789,0.98252)  (1.7037,0.86204)  (2.7565,1.4832)  (2.4196,1.3376)  (2.2203,1.2075)  (2.0871,1.1258)  (2.097,1.0708)  (1.9667,0.97816)  (2.0266,1.2036)  (1.9164,1.1208)  (1.7929,1.0278)  (1.7689,0.97687)  (1.7626,0.93469)  (1.573,0.84243)  (2.0657,1.2249)  (1.9966,1.1482)  (1.9377,1.0883)  (1.9011,1.051)  (1.9401,1.0061)  (1.8569,0.92732) 
             };
             \addlegendentry{$\hat{E}_{\textrm{cross,AV1}}$}

         \addplot[only marks,
            color=black,
            fill=red,
             mark=*,
             mark size=1.5pt,
             line width=0.25pt,
             ]
             coordinates {
         (2.4181,2.4489)  (2.1565,2.1868)  (1.9409,1.957)  (1.7855,1.8088)  (1.6651,1.6988)  (1.5253,1.5615)  (2.5064,2.4481)  (2.2606,2.2285)  (2.0983,2.0616)  (2.0061,1.9445)  (1.9475,1.8617)  (1.8409,1.7516)  (2.4352,2.4693)  (2.1799,2.1155)  (1.9793,1.9552)  (1.8886,1.7821)  (1.7944,1.6727)  (1.6857,1.5799)  (2.3413,2.308)  (2.2033,2.1522)  (2.1205,2.0353)  (2.0328,1.9558)  (1.9907,1.8796)  (1.9286,1.7601)  (2.4148,2.5454)  (2.2124,2.3866)  (2.0832,2.1991)  (1.9571,2.0241)  (1.8458,1.8663)  (1.7332,1.6843)  (2.51,2.5946)  (2.3559,2.4272)  (2.1595,2.2233)  (2.0361,2.0581)  (2.0077,1.9604)  (1.9167,1.7633)  (2.9964,3.1187)  (2.6009,2.779)  (2.336,2.5324)  (2.1125,2.2022)  (1.9418,1.9646)  (1.6515,1.6763)  (2.7556,2.7516)  (2.4757,2.5151)  (2.3033,2.3256)  (2.2323,2.1887)  (2.1643,2.087)  (2.0679,1.9334)  (7.2925,7.7466)  (6.5816,6.7749)  (6.4268,6.4443)  (6.2958,6.2216)  (6.2511,5.9875)  (6.1906,5.8188)  (1.2042,1.2101)  (1.0656,1.1399)  (1.0091,1.0873)  (0.94763,1.04)  (0.91636,0.99161)  (0.90242,0.91475)  (1.1874,1.1846)  (1.1169,1.138)  (1.1072,1.098)  (1.0449,1.0552)  (1.0595,1.0115)  (1.0614,0.94449)  (2.9372,2.8578)  (2.4234,2.3105)  (2.0592,1.9601)  (1.7929,1.719)  (1.6357,1.543)  (1.4263,1.4052)  (2.8434,2.6828)  (2.4244,2.2434)  (2.1395,1.9786)  (1.9841,1.8258)  (1.8503,1.7068)  (1.9067,1.5958)  (6.6786,7.0355)  (6.5142,6.6943)  (6.4366,6.4152)  (6.3783,6.2325)  (6.3411,6.1034)  (6.312,5.9247)  (4.2826,4.0886)  (3.65,3.4748)  (3.084,3.1011)  (2.5074,2.743)  (2.3107,2.4009)  (1.9312,1.8842)  (3.9526,3.7625)  (3.2386,3.2912)  (2.8237,2.9876)  (2.4557,2.6798)  (2.2751,2.3858)  (2.0577,2.0723)  (3.8094,3.9096)  (2.5904,2.5593)  (1.9565,1.8743)  (1.5806,1.6437)  (1.5419,1.5826)  (1.4493,1.4619)  (3.0858,3.168)  (2.2354,2.2234)  (1.9183,1.8586)  (1.8504,1.7979)  (1.8707,1.7899)  (1.8592,1.731)  (1.4238,1.3009)  (1.224,1.1851)  (1.1444,1.1498)  (1.0706,1.0987)  (1.0416,1.0352)  (0.95897,0.94272)  (1.4037,1.2724)  (1.2628,1.2076)  (1.1968,1.1641)  (1.1419,1.1255)  (1.1172,1.0819)  (1.0937,1.0182)  (5.1282,4.4006)  (4.2621,3.97)  (3.271,3.3526)  (2.5748,2.6681)  (2.1786,2.1116)  (1.7834,1.6919)  (4.4683,4.0511)  (3.7192,3.6673)  (3.0992,3.091)  (2.4815,2.4518)  (2.1683,2.1029)  (1.9895,1.8769)  (7.1877,7.6861)  (6.864,7.1541)  (6.6859,6.8446)  (6.5657,6.5871)  (6.4867,6.3691)  (6.4362,6.1899)  (2.764,3.108)  (2.3152,2.3511)  (2.0367,1.9578)  (1.8711,1.783)  (1.779,1.6931)  (1.7053,1.634)  (2.4937,2.6311)  (2.1674,2.2218)  (2.0855,2.1103)  (2.0118,2.0139)  (1.9562,1.9272)  (1.893,1.8141)  (2.0247,1.9014)  (1.8942,1.8499)  (1.763,1.8134)  (1.6263,1.737)  (1.582,1.6723)  (1.5097,1.5737)  (2.2494,2.0138)  (2.1678,1.9792)  (2.042,1.9216)  (1.9431,1.876)  (1.8672,1.8088)  (1.7724,1.7052)  (1.1483,1.1497)  (1.0218,1.0397)  (0.96807,0.9761)  (0.93673,0.93613)  (0.90121,0.91932)  (0.89402,0.90052)  (1.1086,1.1046)  (1.0305,1.0319)  (1.0175,1.0011)  (1.0484,0.9741)  (1.0292,0.95326)  (1.0392,0.93625)  (1.0939,1.116)  (0.98737,1.0363)  (0.92704,0.97009)  (0.91706,0.93479)  (0.89658,0.90764)  (0.87051,0.89389)  (1.116,1.1094)  (1.0511,1.0532)  (1.0431,1.0116)  (1.0491,0.97503)  (1.0548,0.9527)  (1.0551,0.93758)  (2.6354,2.7913)  (2.3017,2.5096)  (2.0337,2.241)  (1.8466,2.0064)  (1.7123,1.8405)  (1.561,1.6583)  (2.7092,2.7281)  (2.4574,2.4991)  (2.2515,2.3045)  (2.113,2.1488)  (2.0844,2.0775)  (2.0245,1.9406)  (2.2582,2.1797)  (1.9139,1.8517)  (1.6769,1.6666)  (1.5679,1.5813)  (1.5149,1.5392)  (1.4648,1.4711)  (2.0625,2.0121)  (1.9939,1.9209)  (1.9168,1.8486)  (1.8857,1.7834)  (1.8991,1.757)  (1.9225,1.7342)  (3.5577,3.5639)  (3.5263,3.3163)  (3.2445,3.0342)  (2.7807,2.7919)  (2.4395,2.5322)  (1.9492,1.977)  (3.909,3.6433)  (3.4593,3.3792)  (2.9912,3.0722)  (2.7283,2.8177)  (2.352,2.5174)  (2.0103,2.0407)  (3.8066,3.5136)  (3.2364,3.1894)  (3.1305,2.8546)  (2.5944,2.5677)  (2.1749,2.2958)  (1.8005,1.7973)  (3.9736,3.6237)  (3.4482,3.2997)  (3.0448,2.9559)  (2.6819,2.6573)  (2.2152,2.3437)  (1.846,1.8736)  (7.8558,6.599)  (6.5463,6.4974)  (6.1273,6.3703)  (5.9535,6.0874)  (5.9658,5.9425)  (5.8102,5.3773)  (9.3945,9.027)  (8.3153,8.4422)  (7.7041,7.7901)  (7.2678,7.2183)  (7.0128,6.8343)  (6.6807,6.2984)  (5.1203,4.6545)  (2.9661,2.7604)  (2.2553,2.144)  (1.9555,1.8605)  (1.8194,1.706)  (1.5371,1.5132)  (3.7985,3.6112)  (2.2449,2.2159)  (2.0063,1.9607)  (1.9725,1.8659)  (1.9313,1.812)  (1.9081,1.772)  (2.3686,2.493)  (1.9897,2.1885)  (1.8239,1.9688)  (1.7409,1.8245)  (1.6341,1.6952)  (1.5524,1.5805)  (2.1661,2.271)  (2.0367,2.0969)  (1.92,1.9695)  (1.8538,1.8618)  (1.7786,1.7795)  (1.8275,1.6936)  (6.8545,7.0363)  (6.5374,6.6994)  (6.3357,6.339)  (6.2701,6.0474)  (6.1547,5.712)  (6.0083,5.3134)  (2.0628,2.2233)  (1.985,2.0813)  (1.8813,1.9596)  (1.8141,1.8621)  (1.704,1.7636)  (1.6363,1.6427)  (2.1472,2.2308)  (2.0426,2.1212)  (1.9907,2.0304)  (1.9231,1.9499)  (1.9441,1.9189)  (1.9066,1.8349)  (2.9338,2.6199)  (2.4269,2.3499)  (1.9885,2.0962)  (1.7521,1.9239)  (1.6756,1.7997)  (1.5729,1.6989)  (2.7161,2.5714)  (2.3144,2.3575)  (2.0196,2.1368)  (1.8501,1.9874)  (1.7531,1.862)  (1.7457,1.7744)  (1.198,1.2023)  (1.1125,1.1535)  (1.071,1.1136)  (1.0146,1.0766)  (1.0049,1.0275)  (0.97064,0.94087)  (1.2643,1.2563)  (1.1995,1.2059)  (1.1288,1.1599)  (1.0904,1.1155)  (1.0822,1.0672)  (1.049,0.97099)  (3.8463,3.6603)  (3.0861,3.1651)  (2.4294,2.7157)  (2.0405,2.342)  (1.8486,2.0164)  (1.6173,1.6829)  (3.6295,3.446)  (2.7613,3.0047)  (2.3163,2.5834)  (2.1805,2.3386)  (2.0746,2.1619)  (1.9955,1.942)  (1.8388,1.9539)  (1.7928,1.9212)  (1.7692,1.8525)  (1.7371,1.8026)  (1.6377,1.7053)  (1.5312,1.5712)  (1.8775,1.9412)  (1.8665,1.8951)  (1.8534,1.8539)  (1.8121,1.7915)  (1.793,1.7166)  (1.7134,1.5673)  (4.4557,3.5625)  (3.8794,3.3498)  (3.8207,3.083)  (3.2872,2.7796)  (2.6752,2.4079)  (1.9974,1.7776)  (4.5551,3.7323)  (3.9849,3.4698)  (3.4304,3.1373)  (3.1944,2.7809)  (2.505,2.3211)  (1.9544,1.8629)  (2.2223,2.1931)  (1.848,1.9136)  (1.6535,1.7316)  (1.5437,1.6548)  (1.5634,1.636)  (1.5821,1.5877)  (1.9835,1.9703)  (1.7493,1.8051)  (1.6734,1.7141)  (1.6591,1.6956)  (1.7007,1.6888)  (1.7615,1.6741)  (3.1293,3.2999)  (2.9421,3.0603)  (2.5486,2.7448)  (2.3445,2.4392)  (2.1403,2.1231)  (1.7352,1.7101)  (3.312,3.2419)  (2.9997,3.0326)  (2.7033,2.7458)  (2.4445,2.4378)  (2.1919,2.1676)  (2.0021,1.8936)  (7.4207,8.0597)  (6.6621,7.052)  (6.5533,6.6531)  (6.4844,6.3903)  (6.4051,6.1915)  (6.2893,5.9591)  (7.3351,7.2766)  (6.9862,6.8773)  (6.6652,6.4574)  (6.4557,6.0842)  (6.5471,5.9997)  (6.3888,5.7441)  (4.3429,4.2102)  (3.9516,3.8062)  (3.0775,3.3397)  (2.5846,2.9064)  (2.1799,2.4999)  (1.8217,1.9718)  (4.0636,3.9706)  (3.4669,3.62)  (3.0478,3.2216)  (2.563,2.8221)  (2.208,2.3849)  (2.0183,2.0056)  (4.1489,3.9104)  (3.5465,3.4002)  (3.0411,2.8841)  (2.4402,2.3525)  (2.1604,1.9866)  (1.8899,1.6918)  (4.614,3.909)  (3.7383,3.4757)  (2.9828,2.8996)  (2.4884,2.4121)  (2.2385,2.1182)  (2.0112,1.8609)  (2.6736,2.7561)  (2.3117,2.4948)  (2.1112,2.2396)  (2.0281,2.0251)  (1.9364,1.8753)  (1.8251,1.7217)  (2.6453,2.7846)  (2.4046,2.5803)  (2.2338,2.3527)  (2.1626,2.2125)  (2.0825,2.1043)  (1.9451,1.9259)  (1.1422,1.1299)  (1.0122,1.0444)  (0.98306,0.99618)  (0.93098,0.95724)  (0.91866,0.92444)  (0.89199,0.89903)  (1.1835,1.1314)  (1.1067,1.0784)  (1.0649,1.0264)  (1.0578,0.98966)  (1.0642,0.96289)  (1.049,0.94214)  (1.1419,1.1603)  (1.0461,1.0818)  (0.97579,1.0079)  (0.93946,0.95829)  (0.90893,0.92294)  (0.88007,0.89569)  (1.1902,1.1639)  (1.084,1.1061)  (1.0603,1.0452)  (1.0581,0.99748)  (1.0474,0.9695)  (1.0621,0.94009)  (1.4561,1.3661)  (1.2872,1.2771)  (1.1442,1.213)  (1.0275,1.1636)  (0.98921,1.1146)  (0.96258,1.008)  (1.4373,1.3878)  (1.2663,1.303)  (1.1659,1.2327)  (1.1048,1.182)  (1.045,1.1187)  (1.0375,1.0026)  (2.8339,2.9058)  (2.4441,2.6237)  (2.2041,2.3618)  (2.0881,2.1008)  (1.8789,1.8819)  (1.7037,1.652)  (2.7565,2.8376)  (2.4196,2.5597)  (2.2203,2.3114)  (2.0871,2.1554)  (2.097,2.0504)  (1.9667,1.8736)  (2.0266,2.304)  (1.9164,2.1458)  (1.7929,1.9685)  (1.7689,1.8712)  (1.7626,1.7907)  (1.573,1.6146)  (2.0657,2.3446)  (1.9966,2.1982)  (1.9377,2.0838)  (1.9011,2.0127)  (1.9401,1.9269)  (1.8569,1.7766) 
             };
             \addlegendentry{$\hat{E}_{\textrm{veri,AV1}}$}

             \addplot[
             color=black,
              mark=none,
              line width=0.5pt,dashed
              ]
              coordinates {
            (0,0)(10,10)
              };
              \addlegendentry{Ideal}
    \end{groupplot}

    \end{tikzpicture}
    \end{center}
    \vspace{-0.5cm}
    \caption{Evaluation of the cross-codec prediction accuracy for the Perf CTC (a) and Valgrind 13PE (b) model. The vertical axis denotes the energy demand prediction of the HW decoder, and the horizontal axis shows the corresponding measurement. Each marker corresponds to one bitstream of the AV1 set.}
    \label{fig:VerificationPlot}

    \vspace{-0.5cm}
    \end{figure}

In Fig.~\ref{fig:VerificationPlot}, we illustrate two examples of cross-codec energy prediction using the Perf CTC (a) and the Valgrind 13PE model (b). The horizontal axis represents the ground truth energy measurement of the AV1 HW decoder, while the vertical axis shows the cross-codec energy prediction $\hat{E}_{\textrm{cross}}$, trained with HEVC and VP9 (phase 7). Each marker corresponds to one bitstream from the dataset.

The blue markers represent the estimated energy demand $\hat{E}{\textrm{cross}}$ based on GPR with pre-trained energy models. In Fig.~\ref{fig:VerificationPlot}b, the Valgrind 13PE model demonstrates a strong correlation between the HW measurement of codec B and $\hat{E}_{\textrm{cross}}$. In contrast, Fig.~\ref{fig:VerificationPlot}a shows a significantly weaker correlation when using the Perf CTC model.

To quantify the linear relationship between $\hat{E}_{\textrm{cross}}$ and $E_{\textrm{veri}}$ (ground truth energy measurement), we use the Pearson correlation coefficient (PCC), which measures the strength of linear association between two variables \cite{Boslaugh2008}.

We observe that the Valgrind 13PE model consistently achieves PCC values exceeding 0.82 for optimized AV1 decoders. When combining multiple video coding standards for training (phases 4–7), the PCC improves further, reaching values over 0.95. In phase 7, a PCC of 0.99 is achieved, indicating a very strong correlation between $\hat{E}_{\textrm{cross}}$ and $E_{\textrm{veri}}$. However, the Temporal model has consistently PCC values below 0.8. Similarly, the Perf CTC model achieves PCC values below 0.83.

Since correlation coefficients are invariant to linear transformations \cite{LeeRodgers1988}, we propose scaling the predicted energy demand using a first-order linear transformation:
\begin{equation}
    \hat{E}_{\textrm{veri}} = \alpha + \hat{E}_{\textrm{cross}} \cdot \beta,
        \label{Eq:Scaling}
\end{equation}
where $\beta$ scales $\hat{E}_{\textrm{cross}}$ to account for differences in technology node, implementation efficiency, and throughput. The offset $\alpha$ captures static energy variations, which can arise from factors like initialization processes in SW (e.g., drivers). Both parameters, $\alpha$ and $\beta$, are trained using LR to fit the actual energy measurement of codec B.

In \cite{KhernacheBenmoussaBoukhobzaEtAl2021}, the authors evaluated two HW video decoders fabricated with different technology nodes and found that both decoders exhibited similar trends in energy consumption. They also observed that the HW decoder with the smaller technology node had a lower static power demand, likely due to its advanced fabrication process.

This finding is significant because it shows that, despite differences in technology nodes, the overall energy consumption trends of different HW decoders can remain consistent. By employing the linear relationship described in equation \eqref{Eq:Scaling}, these HW architecture differences can be accounted for effectively, ensuring that our model is generalizable across various platforms.

In Fig.\ref{fig:VerificationPlot}, the red markers represent the linearly scaled energy demand $\hat{E}_{\textrm{cross}}$. Ideally, the cross-codec prediction should match the measurement, as shown by the dashed line. In Fig.\ref{fig:VerificationPlot}b, we observe that the estimated and measured energy consumptions have a strong linear correlation when using the Valgrind 13PE model. After applying the linear transformation, $\hat{E}_{\textrm{veri}}$ is close to the dashed ideal line.

During the standardization process, training the parameters $\alpha$ and $\beta$ would not be feasible, as a HW implementation is not yet available. As a result, we cannot evaluate the accuracy of the predicted energy demand $\hat{E}_{\textrm{cross}}$, since the corresponding absolute energy demand of an unknown video decoder implementation is not accessible. Additionally, technological advancements between iterations of video coding standards often lead to significant improvements in technology nodes, which cannot be incorporated into the model at this stage.

However, despite these limitations, we can still evaluate the relative influence of processing complexities on the energy efficiency of an unknown HW decoder implementation. This approach allows for the assessment of the impact of new SW complexities, as these are reported and evaluated in the same manner.

\subsection{Verification of Cross-Codec Prediction}
\noindent
As shown in the right box of Fig.~\ref{fig:SolutionOverview}, our verification process involves predicting the HW decoding energy demand for a coding standard that was not included in the model’s training set. To assess the accuracy of these predictions, we apply a linear transformation to compare the predicted energy demand from the GPR ($\hat{E}_{\textrm{cross}}$) with the actual energy measurement of the HW decoder ($E_{\textrm{veri}}$). For example, we could train the energy models using HW decoding measurements and SW decoding profiling from HEVC and VP9, then apply these models to predict the HW energy demand based on SW profiling of AV1.

Table~\ref{tab:VerificationSingleGPR} presents the results of this verification process. The first column lists the video coding standards used for the training of the GPR models, while the second column specifies the SW decoder and standard used for verification. Additionally, we distinguish between reference and optimized SW decoders.

We begin by analyzing cases where a single video coding standard is used for training, corresponding to phases 1-3. When training with AVC (Phase 1), we observe that the MAPE ($\overline{\varepsilon}$) is significantly higher than \pro{10} when predicting the energy demand for AV1. This suggests that AVC is not suitable for effectively cross-predicting the HW decoder energy demand for AV1.

However, the accuracy of the cross-codec prediction improves substantially when HEVC (Phase 2) or VP9 (Phase 3) replaces AVC in training. For the optimized decoders, the Valgrind 13PE model shows an error of \pro{11.85} when trained on VP9 and \pro{8.04} when trained on HEVC. In contrast, the Temporal model has an $\overline{\varepsilon}$ of over \pro{20} on average, and the Perf CTC model shows even higher errors compared to the Temporal model. This indicates that the Valgrind 13PE model outperforms existing methods in predicting the energy demand of an unknown HW decoder implementation, surpassing both SW decoding time and the AOM CTC metrics.

The higher $\overline{\varepsilon}$ observed when training with AVC, compared to training with HEVC or VP9, highlights the significant impact of the evolution of video coding standards on the accuracy of cross-codec predictions. Since AVC was standardized in 2003, it is about a decade older than VP9 and HEVC. Over this period, computational capabilities have increased significantly, allowing for the integration of more advanced coding tools and complex partitioning schemes in newer video coding standards. This evolution has led to the development of more versatile HW decoder implementations, which in turn affects the accuracy of cross-codec predictions.

\begin{table}[!t]

    \def\arraystretch{1.1}
  \centering
  \caption{Analysis of the MAPE for the verification of the cross-codec prediction. The lowest $\overline{\varepsilon}$ for each decoder is given in bold.}
    \label{tab:VerificationSingleGPR}
  \begin{tabular}{c | c | r  | r | r}
    \multicolumn{1}{c|}{}  & \multicolumn{1}{c|}{SW Prof.}  & \multicolumn{1}{c|}{}	& \multicolumn{1}{c|}{Perf}	& \multicolumn{1}{c}{Valgrind} \\
    \multicolumn{1}{c|}{Training}  & \multicolumn{1}{c|}{Verification} & \multicolumn{1}{c|}{Temp.}	& \multicolumn{1}{c|}{ CTC} & \multicolumn{1}{c}{13PE} \\
  \hline \hline
  \multirow{2}{*}{\shortstack[c]{Phase 1\\ AVC}}  
  & AV1 Ref.  & \pros{20.50} & \pros{23.05} & \textbf{\pros{20.11}}  \\ \cdashline{2-5} 
  & AV1 Opt.   & \pros{22.18} & \pros{22.26} &  \textbf{\pros{19.00}} \\ \hline
  \multirow{2}{*}{\shortstack[c]{Phase 2\\ HEVC}}  
  & AV1 Ref.  & \pros{19.48} & \pros{28.10} &  \textbf{\pros{10.17}} \\ \cdashline{2-5} 
  & AV1 Opt.   & \pros{22.08} & \pros{24.81} & \textbf{\pros{8.04}}    \\ \hline
  \multirow{2}{*}{\shortstack[c]{Phase 3\\ VP9}}   
  & AV1 Ref.  & \pros{20.24} & \pros{21.51} & \textbf{\pros{9.18}}    \\ \cdashline{2-5} 
  & AV1 Opt.   & \pros{19.83} & \pros{16.25} &  \textbf{\pros{11.85}}   \\ \hline \hline 
  \multirow{2}{*}{\shortstack[c]{Phase 4\\ AVC,HEVC}}                    
  & AV1 Ref. & \pros{18.64} & \pros{24.98} &  \textbf{\pros{14.55}} \\ \cdashline{2-5} 
  & AV1 Opt.  & \pros{22.77} & \pros{24.26} &  \textbf{\pros{8.24}}    \\ \hline 
  \multirow{2}{*}{\shortstack[c]{Phase 5\\ AVC,VP9}} 
  & AV1 Ref. & \pros{27.26} & \pros{31.82} & \textbf{\pros{14.95}} \\ \cdashline{2-5} 
  & AV1 Opt.  & \pros{20.29} & \pros{20.04} & \textbf{\pros{9.34}} \\ \hline
  \multirow{2}{*}{\shortstack[c]{Phase 6\\ \footnotesize AVC,HEVC,VP9\!\!}}      
  & AV1 Ref. & \pros{22.86} & \pros{32.46} & \textbf{\pros{13.30}} \\ \cdashline{2-5}  
  & AV1 Opt.   & \pros{21.07} & \pros{23.43} & \textbf{\pros{7.07}} \\ \hline 
  \multirow{2}{*}{\shortstack[c]{Phase 7\\ HEVC,VP9}}       
  & AV1 Ref. & \pros{22.49} & \pros{28.90} & \textbf{\pros{10.43}}\\ \cdashline{2-5} 
  & AV1 Opt.   & \pros{21.40} & \pros{23.03} & \textbf{\pros{4.54}}\\ 
  
  \end{tabular}
    
  \vspace{-0.5cm}
  \end{table}

To further enhance cross-codec prediction, we combine multiple video coding standards for training. Our objective is to accurately predict the AV1 decoding energy demand. By merging AVC and HEVC (Phase 4), we achieve an $\overline{\varepsilon}$ of \pro{8.24} using the Valgrind 13PE model for the optimized decoders. Similar results are obtained when combining AVC and VP9 (Phase 5). Training with a combination of AVC, HEVC, and VP9 (Phase 6) leads to an $\overline{\varepsilon}$ of \pro{7.07} for the optimized AV1 decoder using the Valgrind 13PE model.

Finally, we evaluate the effectiveness of training the models with VP9 and HEVC (Phase 7), which results in a significant improvement in $\overline{\varepsilon}$. For the Valgrind 13PE model, we achieve an $\overline{\varepsilon}$ of \pro{4.54}, which is substantially lower compared to the Perf CTC model, where $\overline{\varepsilon}$ is \pro{23.03}. The results of this evaluation are presented in Fig.~\ref{fig:VerificationPlot}.

In Fig.\ref{fig:VerificationPlot}a, we present the cross-codec prediction using the Perf CTC model. This figure shows a broad distribution of energy predictions with high estimation errors, indicating that linear transformation alone cannot sufficiently address these inaccuracies. In contrast, Fig.\ref{fig:VerificationPlot}b shows the results for the Valgrind 13PE model, where applying a linear transformation leads to a significant improvement in the accuracy of HW decoding energy demand predictions.

This considerable enhancement in prediction accuracy suggests that the Valgrind 13PE model is highly beneficial for use in the standardization process. It provides a more reliable estimation of HW decoding energy demands for future HW implementations.

\section{Application in Standardization}
\label{sec:Application}

\begin{figure}[!t]
    \tikzset{box/.style={draw, rectangle, rounded corners=5pt, thick, node distance=2cm, text width=1.75cm,line width=0.03cm,text centered, minimum height=1.5cm}}
    \tikzset{line/.style={draw, line width=0.03cm, -latex}}
  
      \begin{tikzpicture}
  
        \tikzstyle{every node}=[font=\small]
        \node [box] (gprt) {GPR\\Training};

        \node [box, above =0.25cm  of gprt] (swp_av1) {SW Profiling\\AV1};
        \node [box,  right=1.75cm of swp_av1] (gpre_av1) {GPR\\Prediction};

        \node [box, below =0.25cm of gprt] (swp_avm) {SW Profiling\\AVM};
        \node [box,  right=1.75cm of swp_avm] (gpre_avm) {GPR\\Prediction};
  
        \node [box, below right=0.25cm and 0.75cm of gpre_av1] (rehwed) {REHWED};

        \path [line] (gprt) -| (gpre_av1) node[midway, above] {};
        \path [line] (gprt) -| (gpre_avm) node[midway, above left = 0cm and 1.25cm ] {Model};

       \path [line] (swp_av1) -- (gpre_av1) node[midway, above] { $x_{\textrm{PE,AV1}}$};
       \path [line] (swp_avm) -- (gpre_avm) node[midway, below] { $x_{\textrm{PE,AVM}}$};
  
       \path [line] (gpre_av1) -| (rehwed) node[midway, above] {$\hat{E}_{\textrm{cross,AV1}}$};
       \path [line] (gpre_avm) -| (rehwed) node[midway, below] {$\hat{E}_{\textrm{cross,AVM}}$};

    \end{tikzpicture}
  
  \caption{Evaluation of REHWED score using two SW decoders.}
  \label{fig:REHWED}
  \end{figure}
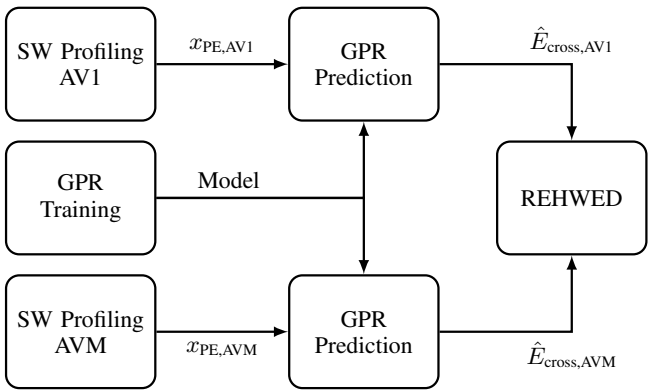

In this section, we introduce the metric Relative Expected HW Energy Demand (REHWED), which utilizes the cross-codec prediction framework to evaluate the energy demand of future decoder implementations during the standardization process. This framework enables us to assess the AVM’s goal of keeping decoder complexity below a target threshold of \pro{200}, relative to AV1~\cite{GoogleAVM}.

To calculate REHWED, we compare the expected energy demand of a future HW video decoder implementation $\hat{E}_{\textrm{cross,test}}$ (e.g., AVM), with an anchor HW video decoder $\hat{E}_{\textrm{cross,anchor}}$ (e.g., AV1), using the following formula:
\begin{equation}
    \textrm{REHWED} = \frac{1}{N} \sum_{i=1}^{N} \frac{\hat{E}_{\textrm{cross,test,}i}}{\hat{E}_{\textrm{cross,anchor,}i}}.
    \label{Eq:REHWED}
\end{equation}

Here, both energy demand predictions are derived using the Valgrind 13PE model, trained on optimized HEVC, VP9, and AV1 software decoders. The ratio of test to anchor is averaged across all bitstreams $i$ in the dataset, consisting of $N$ bitstreams. A pretrained model and implementation for the REHWED metric are provided in~\cite{Rehwed}.

In Fig.~\ref{fig:REHWED}, the evaluation of the REHWED metric is shown. First, SW profiling of AV1 and AVM is conducted, and the corresponding PEs, $x_{\textrm{PE,AV1}}$ and $x_{\textrm{PE,AVM}}$, are used in the GPR prediction to estimate the corresponding HW energy demands, $\hat{E}{\textrm{cross,AV1}}$ and $\hat{E}{\textrm{cross,AVM}}$, respectively. Finally, the ratio of these estimates is compared to calculate the REHWED score.

In Table~\ref{tab:research}, we evaluate the relative differences between the AVM software versions research-alt-v1 (AV1), research-v5.0.0 (v5), research-v6.0.0 (v6), and research-v7.0.0 (v7) in terms of SW decoding time (RSWDT), SW decoding energy demand (RSWED), and REHWED. Note that research-alt-v1 corresponds to AV1 based on the AVM software stack.

We observe in Table~\ref{tab:research} that RSWDT and RSWED exhibit similar values. This aligns with our findings in Section~\ref{sec:SoftwareModeling}, which demonstrated that SW decoding time is a reliable predictor of SW decoding energy demand.

When comparing AV1 and AVM v5, we note that the SW decoding time for v5 increases by \pro{391.83}, while the REHWED for this test reaches \pro{239.51}. This suggests that the expected HW energy demand has more than doubled when compared to AV1. Consequently, this exceeds AVM’s set decoder complexity limit, with an additional energy demand of \pro{139.51} over AV1.

For AVM v7, both RSWDT and RSWED increase significantly by \pro{513.61} and \pro{519.19}, respectively, relative to AV1. Nonetheless, the REHWED for AVM v7 decreases from \pro{241.48} (in v6) to \pro{228.47}.

This reduction in HW energy demand aligns with AOM’s goals for AVM v7, which aims to reduce HW complexity by simplifying and modifying certain coding tools. In summary, while the transition from AV1 to AVM generally results in a notable increase in expected HW decoding energy demand, the changes implemented in AVM v7 help to mitigate this rise, reflecting a balance between improving compression performance and managing energy efficiency.

Based on the presented results, the REHWED metric is used to monitor the expected HW energy demand for future SW anchor releases of AVM, providing key insights into energy efficiency throughout the development process. This proactive approach ensures that future SW iterations align with energy efficiency goals and supports informed decision-making in the development of video coding standards.

\begin{table}[t!]
    \def\arraystretch{1.2}
    \caption{Evaluation of the relative SW decoding time (RSWDT), relative SW energy demand (RSWED), and the proposed REHWED metric for software anchor versions encoded with the RA test configuration.}
    \vspace*{-0.4cm}
    \label{tab:research}
    \begin{center}
    \begin{tabular}{  c | c | c : c : c} 
        Anchor & 	Test &	RSWDT	& RSWED	& REHWED  \\
        \hline
        Research-alt-v1	& Research-v5.0	& \pro{391.83}	& \pro{395.51}	& \pro{239.93} \\
        \hdashline
        Research-alt-v1	& Research-v6.0	& \pro{459.44}	& \pro{459.29}	& \pro{241.48} \\
        \hdashline
        Research-alt-v1	& Research-v7.0	& \pro{513.61}	& \pro{519.19}	& \pro{228.47} \\
    \end{tabular}
    \end{center}
    \vspace*{-0.7cm}
\end{table}

\section{Conclusion}
\label{sec:Conclusion}
\noindent 

In this work, we provided a framework and metric that can be used in standardization to take the expected HW decoding energy demand into consideration.
We show that the processor-events based 13PE model has a MAPE of \pro{1.79} for the estimation of HW decoders and a MAPE of \pro{1.38} for SW decoders. 
Moreover, the developed metric allows to cross-codec predict the energy demand of an unknown HW implementation with a MAPE of \pro{4.54} without using the corresponding HW decoder for training.

The proposed methodology helps standardization bodies to develop future video coding standards with enhanced HW energy efficiency, leading to reduced GHG emissions and longer battery life for mobile devices. 
The provided experiments showed evidence that SW decoders can be used to accurately predict the HW decoder energy demand during the development of a future standard. 
Consequently, the standardization process can use such predictions to select coding tools with lower energy demand, which may also lead to a smaller area.

In future work, we plan to investigate hardware-specific factors, such as DDR bandwidth and on-chip memory consumption, as well as the impact of different technology nodes on prediction accuracy. Furthermore, this methodology could be extended to other HW video encoders or algorithms that provide SW and HW implementations.

\bibliographystyle{IEEEtran}

\begin{thebibliography}{10}
    \providecommand{\url}[1]{#1}
    \csname url@samestyle\endcsname
    \providecommand{\newblock}{\relax}
    \providecommand{\bibinfo}[2]{#2}
    \providecommand{\BIBentrySTDinterwordspacing}{\spaceskip=0pt\relax}
    \providecommand{\BIBentryALTinterwordstretchfactor}{4}
    \providecommand{\BIBentryALTinterwordspacing}{\spaceskip=\fontdimen2\font plus
    \BIBentryALTinterwordstretchfactor\fontdimen3\font minus
      \fontdimen4\font\relax}
    \providecommand{\BIBforeignlanguage}[2]{{%
    \expandafter\ifx\csname l@#1\endcsname\relax
    \typeout{** WARNING: IEEEtran.bst: No hyphenation pattern has been}%
    \typeout{** loaded for the language `#1'. Using the pattern for}%
    \typeout{** the default language instead.}%
    \else
    \language=\csname l@#1\endcsname
    \fi
    #2}}
    \providecommand{\BIBdecl}{\relax}
    \BIBdecl
    
    \bibitem{Ericson2023}
    \BIBentryALTinterwordspacing
    {Ericson}. (2023, Jun.) Ericson mobility report. [Online]. Available:
      \url{https://www.ericsson.com/49dd9d/assets/local/reports-papers/mobility-report/documents/2023/ericsson-mobility-report-june-2023.pdf}
    \BIBentrySTDinterwordspacing
    
    \bibitem{Efoui-Hess2019}
    \BIBentryALTinterwordspacing
    M.~Efoui-Hess. (2019, Jul.) Climate crisis: The unsustainable use of online
      video. {T}he practical case for digital sobriety. [Online]. Available:
      \url{https://theshiftproject.org/en/article/unsustainable-use-online-video/}
    \BIBentrySTDinterwordspacing
    
    \bibitem{Herglotz2023c}
    C.~Herglotz, M.~Kränzler, R.~Schober, and A.~Kaup, ``Sweet streams are made of
      this: The system engineer’s view on energy efficiency in video
      communications,'' \emph{IEEE Circ. and Sys. Magazine}, vol.~23, no.~1, pp.
      57--77, April 2023.
    
    \bibitem{Herglotz19a}
    C.~Herglotz, S.~Coulombe, S.~Vakili, and A.~Kaup, ``Power modeling for virtual
      reality video playback applications,'' in \emph{Proc. IEEE Int. Symp. on
      Consumer Tech. (ISCT)}, Ancona, Italy, Jun. 2019.
    
    \bibitem{avm}
    \BIBentryALTinterwordspacing
    Google. {AVM Codec (v3.0)}. Accessed 2022-07. [Online]. Available:
      \url{https://gitlab.com/AOMediaCodec/avm}
    \BIBentrySTDinterwordspacing
    
    \bibitem{Kraenzler2020MMSP}
    M.~{Kr\"anzler}, C.~{Herglotz}, and A.~{Kaup}, ``A comparative analysis of the
      time and energy demand of versatile video coding and high efficiency video
      coding reference decoders,'' in \emph{Proc. IEEE Int. Workshop on Multimedia
      Signal Processing (MMSP)}, Tampere, Finland, Sep. 2020.
    
    \bibitem{Kranzler_2024}
    M.~{Kr\"anzler}, A.~Kaup, and C.~Herglotz, ``A comprehensive review of software
      and hardware energy efficiency of video decoders,'' in \emph{Proc. Picture
      Coding Symp. ({PCS})}, Taichung, Taiwan, Jun. 2024.
    
    \bibitem{VCEG-M33}
    G.~Bj{\o}ntegaard, ``Calculation of average {PSNR} differences between {RD}
      curves,'' Austin, TX, USA, {document, VCEG-M33}, Jan. 2001.
    
    \bibitem{Herglotz2023}
    C.~Herglotz \emph{et~al.}, ``The bj{\o}ntegaard bible -- why your way of
      comparing video codecs may be wrong,'' \emph{IEEE Trans. on Image
      Processing}, vol.~33, pp. 987--1001, Jan. 2024.
    
    \bibitem{valgrind}
    \BIBentryALTinterwordspacing
    Valgrind. Accessed 2021-10-01. [Online]. Available: \url{https://valgrind.org/}
    \BIBentrySTDinterwordspacing
    
    \bibitem{Laude_2019}
    T.~Laude, Y.~G. Adhisantoso, J.~Voges, M.~Munderloh, and J.~Ostermann, ``A
      comprehensive video codec comparison,'' \emph{{APSIPA} Trans. on Signal and
      Information Processing}, vol.~8, no.~1, 2019.
    
    \bibitem{Mercat_2021}
    A.~Mercat, A.~Makinen, J.~Sainio, A.~Lemmetti, M.~Viitanen, and J.~Vanne,
      ``Comparative rate-distortion-complexity analysis of {VVC} and {HEVC} video
      codecs,'' \emph{{IEEE} Access}, vol.~9, pp. 67\,813--67\,828, 2021.
    
    \bibitem{Nguyen_2021}
    T.~Nguyen and D.~Marpe, ``Compression efficiency analysis of {AV}1, {VVC}, and
      {HEVC} for random access applications,'' \emph{{APSIPA} Trans. on Signal and
      Information Processing}, vol.~10, no.~1, 2021.
    
    \bibitem{Katsenou_2022}
    A.~Katsenou, J.~Mao, and I.~Mavromatis, ``Energy-rate-quality tradeoffs of
      state-of-the-art video codecs,'' in \emph{Proc. Picture Coding Symp.
      ({PCS})}, San Jose, CA, USA, Dec. 2022.
    
    \bibitem{Correa_2021}
    M.~Correa \emph{et~al.}, ``{AV}1 and {VVC} video codecs: Overview on complexity
      reduction and hardware design,'' \emph{{IEEE} Open Journal of Circ. and
      Sys.}, vol.~2, pp. 564--576, 2021.
    
    \bibitem{Hamidouche_2022}
    W.~Hamidouche, P.~Philippe, S.~A. Fezza, M.~Haddou, F.~Pescador, and D.~Menard,
      ``Hardware-friendly multiple transform selection module for the vvc
      standard,'' \emph{IEEE Trans. on Consumer Electronics}, vol.~68, no.~2, pp.
      96--106, May 2022.
    
    \bibitem{Farhat_2021}
    I.~Farhat, W.~Hamidouche, A.~Grill, D.~Ménard, and O.~Déforges, ``Lightweight
      hardware transform design for the versatile video coding 4k asic decoders,''
      \emph{IEEE Trans. on Consumer Electronics}, vol.~67, no.~4, pp. 329--340,
      Nov. 2021.
    
    \bibitem{Farhat_2022}
    ------, ``Adaptive loop filter hardware design for 4k asic vvc decoders,''
      \emph{IEEE Trans. on Consumer Electronics}, vol.~68, no.~2, pp. 107--118, May
      2022.
    
    \bibitem{Saldanha_2020}
    M.~Saldanha \emph{et~al.}, ``An overview of dedicated hardware designs for
      state-of-the-art {AV}1 and h.266/{VVC} video codecs,'' in \emph{Proc. {IEEE}
      Int. Conf. on Electronics, Circ. and Sys. ({ICECS})}.\hskip 1em plus 0.5em
      minus 0.4em\relax {IEEE}, Nov. 2020.
    
    \bibitem{Viitanen2022}
    M.~Viitanen, J.~Sainio, A.~Mercat, A.~Lemmetti, and J.~Vanne, ``From hevc to
      vvc: The first development steps of a practical intra video encoder,''
      \emph{IEEE Trans. on Consumer Electronics}, vol.~68, no.~2, pp. 139--148, May
      2022.
    
    \bibitem{Wieckowski2022}
    A.~Wieckowski, J.~Brandenburg, B.~Bross, and D.~Marpe, ``Vvc search space
      analysis including an open, optimized implementation,'' \emph{IEEE Trans. on
      Consumer Electronics}, vol.~68, no.~2, pp. 127--138, May 2022.
    
    \bibitem{Herglotz2017}
    C.~{Herglotz} and A.~{Kaup}, ``Video decoding energy estimation using processor
      events,'' in \emph{Proc. IEEE Int. Conf. on Image Processing (ICIP)},
      Beijing, China, Sep. 2017.
    
    \bibitem{MallikarachchiTalagalaH.EtAl2017}
    T.~Mallikarachchi, D.~S. Talagala, H.~K. {Arachchi}, and A.~Fernando, ``A
      feature based complexity model for decoder complexity optimized {HEVC} video
      encoding,'' in \emph{Proc. IEEE Int. Conf. on Consumer Electronics (ICCE)},
      Las Vegas, NV, USA, Jan. 2017.
    
    \bibitem{HerglotzSpringerReichenbachEtAl2018}
    C.~{Herglotz}, D.~{Springer}, M.~{Reichenbach}, B.~{Stabernack}, and A.~{Kaup},
      ``Modeling the energy consumption of the {HEVC} decoding process,''
      \emph{IEEE Trans. on Circ. and Sys. for Video Tech.}, vol.~28, no.~1, pp.
      217--229, Jan. 2018.
    
    \bibitem{Kraenzler2019}
    M.~{Kr\"anzler}, C.~{Herglotz}, and A.~{Kaup}, ``Extending video decoding
      energy models for 360$^\circ$ and {HDR} video formats in {HEVC},'' in
      \emph{Proc. Picture Coding Symp. (PCS)}, Ningbo, China, Nov. 2019.
    
    \bibitem{Kraenzler2020ICIP}
    ------, ``Decoding energy modeling for versatile video coding,'' in \emph{Proc.
      Int. Conf. on Image Processing (ICIP)}, Abu Dhabi, United Arab Emirates, Oct.
      2020.
    
    \bibitem{Herglotz_2018}
    C.~Herglotz and A.~Kaup, ``Decoding energy estimation of an {HEVC} hardware
      decoder,'' in \emph{2018 {IEEE} Int. Symp. on Circ. and Sys. ({ISCAS})},
      Florence, Italy, May 2018.
    
    \bibitem{HerglotzCoulombeVazquezEtAl2020}
    C.~Herglotz, S.~Coulombe, C.~Vazquez, A.~Vakili, A.~Kaup, and J.-C. Grenier,
      ``Power modeling for video streaming applications on mobile devices,''
      \emph{IEEE Access}, vol.~8, pp. 70\,234--70\,244, Apr. 2020.
    
    \bibitem{Kraenzler2023}
    M.~{Kr\"anzler}, A.~{Kaup}, and C.~{Herglotz}, ``Estimating software and
      hardware video decoder energy using software decoder profiling,'' in
      \emph{Proc. 36th SBC/SBMicro/IEEE/ACM Symp. on Integrated Circ. and Sys.
      Design (SBCCI)}, Rio de Janeiro, Brazil, Aug. 2023.
    
    \bibitem{HerglotzHeindelKaup}
    C.~{Herglotz}, A.~{Heindel}, and A.~{Kaup}, ``Decoding-energy-rate-distortion
      optimization for video coding,'' \emph{IEEE Trans. on Circ. and Sys. for
      Video Tech.}, vol.~29, no.~1, pp. 171--182, Jan. 2019.
    
    \bibitem{Kraenzler21}
    M.~Kr\"anzler, C.~Herglotz, and A.~Kaup, ``{Energy Efficient Video Decoding for
      VVC Using a Greedy Strategy Based Design Space Exploration},'' \emph{IEEE
      Trans. on Circ. and Sys. for Video Tech.}, vol.~32, no.~7, pp. 4696--4709,
      Jul. 2022.
    
    \bibitem{Kraenzler2022}
    M.~Kränzler \emph{et~al.}, ``Optimized decoding-energy-aware encoding in
      practical {VVC} implementations,'' in \emph{Proc. Int. Conf. on Image
      Processing (ICIP)}, Bordeaux, France, Oct. 2022.
    
    \bibitem{Kranzler_2022}
    M.~{Kr\"anzler}, A.~Kaup, and C.~Herglotz, ``Advanced design space exploration
      for joint energy and quality optimization for {VVC},'' in \emph{Proc. Picture
      Coding Symp. ({PCS})}, San Jose, CA, USA, Dec. 2022.
    
    \bibitem{Tissier_2019}
    A.~Tissier, A.~Mercat, T.~Amestoy, W.~Hamidouche, J.~Vanne, and D.~Menard,
      ``Complexity reduction opportunities in the future {VVC} intra encoder,'' in
      \emph{Proc. Int. Workshop on Multimedia Signal Processing ({MMSP})}, Kuala
      Lumpur, Malaysia, Sep. 2019.
    
    \bibitem{Choi_2022}
    K.~Choi, T.~V. Le, Y.~Choi, and J.~Y. Lee, ``Low-complexity intra coding in
      versatile video coding,'' \emph{IEEE Trans. on Consumer Electronics},
      vol.~68, no.~2, pp. 119--126, May 2022.
    
    \bibitem{Amirpour_2023}
    H.~Amirpour, V.~V. Menon, S.~Afzal, R.~Prodan, and C.~Timmerer, ``Optimizing
      video streaming for sustainability and quality: The role of preset selection
      in per-title encoding,'' in \emph{Proc. Int. Conf. on Multimedia and Expo
      ({ICME})}, Brisbane, Australia, Jul. 2023.
    
    \bibitem{x264}
    \BIBentryALTinterwordspacing
    Videolan. {x264 Encoder}. Accessed 2021-09. [Online]. Available:
      \url{https://code.videolan.org/videolan/x264.git}
    \BIBentrySTDinterwordspacing
    
    \bibitem{JM}
    \BIBentryALTinterwordspacing
    {HHI Fraunhofer}. {JM Decoder (v19.1)}. Accessed 2021-09. [Online]. Available:
      \url{https://vcgit.hhi.fraunhofer.de/jvet/JM}
    \BIBentrySTDinterwordspacing
    
    \bibitem{ffmpeg}
    \BIBentryALTinterwordspacing
    {Fast Forwards MPEG (FFmpeg)}. Accessed 2018-11-14. [Online]. Available:
      \url{http://ffmpeg.org/}
    \BIBentrySTDinterwordspacing
    
    \bibitem{x265}
    \BIBentryALTinterwordspacing
    Videolan. {x265 Encoder (v3.5.1}. Accessed 2021-09. [Online]. Available:
      \url{http://hg.videolan.org/x265}
    \BIBentrySTDinterwordspacing
    
    \bibitem{HM}
    \BIBentryALTinterwordspacing
    {HHI Fraunhofer}. {HM Decoder (v16.23)}. Accessed 2021-09. [Online]. Available:
      \url{https://vcgit.hhi.fraunhofer.de/jvet/HM}
    \BIBentrySTDinterwordspacing
    
    \bibitem{openHEVC}
    \BIBentryALTinterwordspacing
    {openHEVC}. Accessed 2021-02-25. [Online]. Available:
      \url{https://github.com/OpenHEVC/openHEVC}
    \BIBentrySTDinterwordspacing
    
    \bibitem{VVenCSoftware}
    \BIBentryALTinterwordspacing
    {Fraunhofer HHI VVenC Software Repository (v1.7)}. Accessed: Jan 2022.
      [Online]. Available: \url{https://github.com/fraunhoferhhi/vvenc}
    \BIBentrySTDinterwordspacing
    
    \bibitem{VTM}
    \BIBentryALTinterwordspacing
    {Joint Video Exploration Team ({JVET})}. {VVC test model reference software
      (v19.0)}. [Online]. Available:
      \url{https://vcgit.hhi.fraunhofer.de/jvet/VVCSoftware_VTM/}
    \BIBentrySTDinterwordspacing
    
    \bibitem{VVdeCSoftware}
    \BIBentryALTinterwordspacing
    {Fraunhofer HHI VVdeC Software Repository (v1.6)}. Accessed: Jan 2022.
      [Online]. Available: \url{https://github.com/fraunhoferhhi/vvdec}
    \BIBentrySTDinterwordspacing
    
    \bibitem{libvpx}
    \BIBentryALTinterwordspacing
    Google. {libvpx Codec (v1.10)}. Accessed 2021-10. [Online]. Available:
      \url{https://chromium.googlesource.com/webm/libvpx/}
    \BIBentrySTDinterwordspacing
    
    \bibitem{libaom}
    \BIBentryALTinterwordspacing
    A.~for Open~Media. {libaom Codec (v3.3)}. Accessed 2022-03. [Online].
      Available: \url{https://aomedia.googlesource.com/aom/}
    \BIBentrySTDinterwordspacing
    
    \bibitem{dav1d}
    \BIBentryALTinterwordspacing
    DAV1D. {DAV1D Software (v1.0)}. Accessed 2022-03. [Online]. Available:
      \url{https://code.videolan.org/videolan/dav1d}
    \BIBentrySTDinterwordspacing
    
    \bibitem{CWGB005oV1}
    X.~Zhao, Z.~Lei, A.~Norkin, T.~Daede, and A.~Tourapis, ``{AV2} common test
      conditions v1.0,'' document, CWG-B005o v1, Jan. 2021.
    
    \bibitem{DavidGorbatovHanebutteEtAl2010}
    H.~David, E.~Gorbatov, U.~R. Hanebutte, R.~Khanna, and C.~Le, ``{RAPL}: Memory
      power estimation and capping,'' in \emph{Proc. ACM/IEEE Int. Symp. on
      Low-Power Electronics and Design (ISLPED)}, Austin, TX, USA, Aug. 2010.
    
    \bibitem{Katsenou2024}
    A.~Katsenou, X.~Wang, D.~Schien, and D.~Bull, ``Comparative study of hardware
      and software power measurements in video compression,'' in \emph{Proc.
      Picture Coding Symp. (PCS)}, Taichung, Taiwan, June 2024.
    
    \bibitem{Rock5B}
    \BIBentryALTinterwordspacing
    Radxa. {Rock 5B Hardware Details}. Accessed 2023-11. [Online]. Available:
      \url{https://wiki.radxa.com/Rock5/hardware/5b}
    \BIBentrySTDinterwordspacing
    
    \bibitem{Izenman2013}
    A.~J. Izenman, \emph{Modern Multivariate Statistical Techniques: Regression,
      Classification, and Manifold Learning}.\hskip 1em plus 0.5em minus
      0.4em\relax Springer, 2013.
    
    \bibitem{ColemanLi1996}
    T.~F. {Coleman} and Y.~{Li}, ``An interior trust region approach for nonlinear
      minimization subject to bounds,'' \emph{SIAM Journal on optimization},
      vol.~6, no.~2, pp. 418--445, May 1996.
    
    \bibitem{Rasmussen2006}
    C.~E. Rasmussen and C.~K.~I. Williams, \emph{Gaussian Processes for Machine
      Learning}.\hskip 1em plus 0.5em minus 0.4em\relax Cambridge, MA, USA: MIT
      Press, 2006.
    
    \bibitem{Matlab2023}
    \BIBentryALTinterwordspacing
    Matlab. Gaussian process regression models. Accessed 2023-11. [Online].
      Available:
      \url{https://de.mathworks.com/help/stats/gaussian-process-regression-models.html}
    \BIBentrySTDinterwordspacing
    
    \bibitem{Hastie2009}
    T.~Hastie, R.~Tibshirani, and J.~Friedman, \emph{The Elements of Statistical
      Learning: Data Mining, Inference, and Prediction, Second Edition}.\hskip 1em
      plus 0.5em minus 0.4em\relax Springer New York, NY, USA, 2009.
    
    \bibitem{Boslaugh2008}
    P.~A. Watters and S.~Boslaugh, \emph{Statistics in a nutshell: A desktop quick
      reference}.\hskip 1em plus 0.5em minus 0.4em\relax Sebastopol, CA, USA:
      O'Reilly, 2008.
    
    \bibitem{LeeRodgers1988}
    J.~Lee~Rodgers and W.~A. Nicewander, ``Thirteen ways to look at the correlation
      coefficient,'' \emph{The American Statistician}, vol.~42, no.~1, pp. 59--66.
    
    \bibitem{KhernacheBenmoussaBoukhobzaEtAl2021}
    M.~B.~A. Khernache, Y.~Benmoussa, J.~Boukhobza, and D.~Menard, ``{HEVC}
      hardware vs software decoding: An objective energy consumption analysis and
      comparison,'' \emph{Journal of Sys. Architecture}, vol. 115, p. 102004, May
      2021.
    
    \bibitem{GoogleAVM}
    I.~S. Chong, J.~Young, S.~Li, C.~McCullough, S.~Vitvitskyy, and V.-M. Rautio,
      ``Av1/avm development at google,'' in \emph{Proc. SPIE 13137, Applications of
      Digital Image Processing XLII}, San Diego, CA, United States, Sep. 2024.
    
    \bibitem{Rehwed}
    \BIBentryALTinterwordspacing
    M.~{Kr\"anzler}, C.~{Herglotz}, and A.~{Kaup}. {HW Energy Estimation
      Framework}. Accessed 2024-11. [Online]. Available:
      \url{https://github.com/FAU-LMS/hw-energy-estimation}
    \BIBentrySTDinterwordspacing
    
    \end{thebibliography}


\end{document}